%% file: main.tex
\documentclass[reprint,showkeys]{revtex4-2}
\usepackage{float}
\usepackage{amsmath,amsthm,amsfonts,amssymb,amscd}
\usepackage{graphicx}
\usepackage{bm} 
\DeclareMathOperator{\diag}{diag}

\newcommand{\beginsupplement}{%
        \setcounter{table}{0}
        \renewcommand{\thetable}{S\arabic{table}}%
        \setcounter{figure}{0}
        \renewcommand{\thefigure}{S\arabic{figure}}%
     }

\begin{document}

\title{Transmission grid stability with large interregional power flows}

\author{Mar\'{i}a Mart\'{i}nez-Barbeito}
\email{Corresponding author: maria@ifisc.uib-csic.es (she/her/hers)}
\author{Dami\`a Gomila}
\author{Pere Colet}
\affiliation{Instituto de F\'{i}sica Interdisciplinar y Sistemas Complejos (IFISC, UIB-CSIC), Campus Universitat de les Illes Balears E-07122, Palma de Mallorca, Spain}
\author{Julian Fritzsch}
\author{Philippe Jacquod}
\affiliation{Department of Quantum Matter Physics, University of Geneva, CH-1211 Geneva, Switzerland}
\affiliation{School of Engineering, University of Applied Sciences of Western Switzerland HES-SO, CH-1950 Sion, Switzerland}

\date{\today}

\begin{abstract} 
We propose a general methodology for identifying critical lines in the long-distance transmission of power across large electric grids. When the system is pushed to its operational limit, for instance by large power imbalances such as those generated by high penetration of variable renewable energy sources, the network gets destabilized and loses synchrony. We investigate a model of the synchronous AC grid of continental Europe under tunable large interregional power flows. When those flows exceed some critical value, we find that instabilities emerge due to topological constraints. We
identify two different scenarios triggering these instabilities. 
In the first one, specific sets of lines reach
their maximal load simultaneously, causing the grid to split into two desynchronized zones. 
In the second one, one or few lines become overloaded, which eventually renders one Lyapunov exponent positive. The first scenario is obviously less generic and we develop a numerical approach to force the splitting of the AC grid into 
disconnected areas. Remarkably, the critical lines identified in this way match those that triggered the separation of the synchronous grid of continental Europe in two instances in 2021.  
We further discuss  how the modes of the system provide information on which areas are more susceptible to lose synchrony with each other.
\end{abstract} 


\maketitle

\section{Introduction} 

The steep increase in atmospheric and oceanic temperatures of the past decades is a direct consequence of anthropogenic emissions of greenhouse gases~\cite{Fle98,IPCC19}, with carbon dioxide (CO$_{2}$) being a major concern due to its high emission volume and century-long atmospheric lifetime. Decarbonization -- suppressing CO$_{2}$ emissions from human activities -- is therefore crucial for climate change mitigation.   Many countries aim for climate neutrality by 2050 ~\cite{europeangreendeal,californiacarbonneutral,usacarbonneutral,chinacarbonneutral} as part of an international effort to combat climate change \cite{parisagreement}.

Achieving this goal requires substantial changes across various sectors of the economy and, in particular, transforming the energy sector, which is the main contributor to CO$_{2}$ emissions~\cite{europeancommissionCO2EmissionsAll2022}. A shift to low-carbon energy sources is therefore essential, with Variable Renewable Energy Sources (VRES) expected to increase massively worldwide in upcoming years. Within Europe, for instance, electricity generation is expected to rely on an optimal combination of wind and solar power \cite{heide2010seasonal}.

The integration of a large fraction of VRES in the power grid poses significant challenges \cite{anees2012grid, basit2020limitations}. In particular, the transition from synchronous generators to a more electronics-based power system raises power quality issues \cite{liang2016emerging,kumar2016grid,denholm2020inertia,makolo2021role}. Various alternatives have been put forward to address these difficulties \cite{behabtu2020review,zhong2010synchronverters,fang2018inertia,fritzsch23}.

Other key aspects that need to be considered are environmental and  geographical factors restricting the deployment of VRES. Some regions possess greater renewable potential than others, hence they are likely to contribute more to the VRES generation share. In Europe for instance, Germany, Spain, France, and Italy, have the highest wind turbine capacity plans \cite{monforti2016synchronous}. VRES are both fluctuating and undispatchable, which will inevitably lead to local excess or lack of power supply. These can be significantly mitigated by large-scale power transfers over distant areas with different power mixes and varying meteorological conditions. Consequently, in parallel to decarbonization and the deployment of large quantities of VRES, there is an urgent need for strengthening grid infrastructures, to help them support large interregional power flows across national borders and between different system operators \cite{zappa2019100}. Grid upgrades are costly and it is therefore crucial to develop methods to identify high-priority upgrades.

Today's large-scale high-voltage transmission grids have evolved from the interconnection of national and regional subsystems over decades. They were  not originally designed to support large cross-border or interregional trades \cite{gridhistory}. Furthermore, Transmission System Operators (TSOs) traditionally look after their own control area and exchange little real-time information with their counterpart in other areas of the same interconnected power grid. This makes the response to contingencies slower and inadequate, and the grid more prone to blackouts \cite{UCTE2004, bialek2007has, li2007analysis, van2010transnational, fotis2023risks}. While the existing infrastructure and policies have improved in recent years \cite{eurelectric2020}, the expected cross-border exchanges may exceed the capacity of the current system design \cite{rodriguez2014transmission}, which will push the system close to its limits.

It is therefore crucial to better understand how the upcoming large amounts of VRES will impact the legacy-power-system on which they will be deployed. This is the main focus of this article.  Our goal is to analyze grid stability in scenarios with large interregional power flows and identify which are the most vulnerable areas, and which are the most sensitive lines to these large interregional flows.

The paper is structured as follows. Section \ref{sec:model} describes the dynamical model. Section \ref{sec:European_grid} describes the European grid model on which we will focus and the methodology. In Sections \ref{sec:north-south_central-east} and \ref{sec:DC_trick}, we present our results. Section \ref{sec:DC_trick} in particular presents a novel methodology for identifying cutsets, i.e. sets of lines that may lead to grid separation into disconnected areas. Section \ref{sec:conclusions} summarises the main discussion points and gives concluding remarks. We stress that, while our numerical investigations focus on
a model of the transmission grid of continental Europe, our conclusions apply to other large-scale electric power grids.

\section{Model} \label{sec:model}

In the lossless line approximation, which is justified and standardly used when considering ultra-high voltage transmission grids~\cite{Machowski2008}, and assuming constant voltage amplitude, the dynamics of a transmission grid can be described by the swing equation
\begin{equation} \label{eq:oscillator}
    m_{i} \ddot{\theta}_{i} + d_{i} \dot{\theta}_{i} = p_{i} - \sum_{j} b_{ij} \sin (\theta_{i} -\theta_{j}), 
\end{equation}
where $\theta_{i}$ is the voltage phase of bus $i$, for $i=1,...,N$, $m_{i}$ and $d_{i}$ are its inertia and damping coefficients, $p_i>0$ is the power generated and $p_i<0$ the power consumed,
and $b_{ij}=B_{ij}V_{ij}^{2}$ is the theoretical maximum power flow for the transmission line between nodes $i$ and $j$ determined by its susceptance $B_{ij}$ (imaginary part of the admittance) and voltage $V_{ij}$. As in Ref.~\cite{pagnier2019inertia}, generator buses have inertia, $m_i>0$, $i=1,..., N_{\rm G}$, while load buses are considered inertia-less, i.e. $m_{i}=0$, $i=N_{\rm G}+1, ..., N$.

This is the \textit{structure preserving model} presented by Bergen and Hill in 1981 \cite{bergen1981structure}. Since their early works \cite{hill1982stability}, to more recent studies \cite{song2017network,song2017characterization}, power grid stability has been topic of study for decades \cite{liu2022stability}. In what follows, we revisit the stability analysis done in Refs.~\cite{coletta2016linear,manik2014supply} for the case in which all nodes have inertia. Then, we extend the results to the general case in which there are heterogeneous and inertia-less nodes.

The fixed points of Eq.~(\ref{eq:oscillator}) are given by $\dot{\theta_i^*} =0$ and $p^*_{i} - \sum_{j} b_{ij} \sin (\theta_{i}^* -\theta_{j}^*)=0$. In the case where all nodes have inertia, for  given $\lbrace p_i \rbrace$, the stability of the fixed points is determined by the Jacobian
\begin{equation}
    \bm{J}= \begin{bmatrix}\bm{0} & \mathbb{I}\\
    -\bm{M}^{-1}\bm{L} \;& -\bm{M}^{-1}\bm{D}\end{bmatrix},
\end{equation}
where $\bm{M} = \diag(m_1,\dots, m_N)$, $\bm{D} = \diag(d_1,\dots, d_N)$, and $\bm{L}$ is the weighted network Laplacian defined as
\begin{equation}\label{eq:Laplacian}
    \bm{L}_{ij} =
    \begin{cases}
        -b_{ij}\cos(\theta_i^* - \theta_j^*) & \text{for } i\neq j \\
        \sum_k b_{ik}\cos(\theta_i^* - \theta_k^*) & \text{for } i = j
    \end{cases}.
\end{equation}
The subscript $i$ ($j$) denotes the rows (columns) corresponding to the $i$th ($j$th) node.

The eigenvalues $\Lambda$ of the Jacobian are determined by
\begin{equation}
\begin{vmatrix} 
-\Lambda \mathbb{I} & \mathbb{I}\\
-\bm{M}^{-1}\bm{L}  \;& -\bm{M}^{-1}\bm{D} -\Lambda\mathbb{I}\end{vmatrix} =0,
\label{eq:determinant_inertia}
\end{equation}
hence
\begin{equation}
    \left| \Lambda (\bm{M}^{-1}\bm{D} + \Lambda\mathbb{I}) + \bm{M}^{-1}\bm{L} \right| =0.
\label{eq:determinant_expanded}
\end{equation}
This equation has a simple solution if the damping of the nodes is proportional to their inertia, namely
\begin{equation}
    \frac{d_i}{m_i} =\kappa, \; \;  i=1,\dots,N.
    \label{eq:damping_propto_inertia}
\end{equation}
In this case, Eq.~(\ref{eq:determinant_expanded}) becomes
\begin{equation}
    \left| \Lambda (\kappa +\Lambda)\mathbb{I} + \bm{M}^{-1}\bm{L} \right| =0.
\end{equation}
Defining $\lambda=-\Lambda(\kappa+\Lambda)$, this is the eigenvalue equation for the matrix $\bm{M}^{-1}\bm{L}$. Thus,
the eigenvalues of the Jacobian come in pairs: for each eigenvalue $\lambda_i$ ($i=1,\dots, N$) of $\bm{M}^{-1}\bm{L}$, we have two eigenvalues of the Jacobian given by
\begin{equation}
    \Lambda_{i}^{\pm} = 
    \frac{-\kappa \pm \sqrt{\kappa^2-4\lambda_i }}{2}.
    \label{eq:eigenvalues_proportionality_case}
\end{equation}

In general, however, $\bm{M}^{-1}\bm{D}$ is not proportional to the identity matrix. Therefore, Eq.~(\ref{eq:determinant_expanded}) can not be solved easily. Nevertheless, it is possible to gain some insight from the Jacobian eigenvector equation. The phase, $\vec{u}_j$, and frequency, $\vec{w}_j$, components of the eigenvector $j$ ($j=1,\dots,2N$) fulfill
\begin{equation}
\bm{J}\begin{bmatrix} \vec{u}_j \\ \vec{w}_j    \end{bmatrix} = \Lambda_{j} \begin{bmatrix} \vec{u}_j \\ \vec{w}_j    \end{bmatrix} .
\end{equation}
Thus,
\begin{align}
    &\vec{w}_j = \Lambda_j {\vec u}_j, \\
    &-\bm{L} \vec{u}_j  -\Lambda_j \bm{D} \vec{u}_j = \Lambda^2\bm{M}\vec{u}_j.
    \label{eq:phase_eigenvectors}
\end{align}
Multiplying the left hand side by $\vec{u}_j^\dagger$ and defining 
$\alpha_j=\vec{u}_j^{\dagger} \mathbf{M} \vec{u}_j$, $\beta_j=\vec{u}_j^{\dagger} \mathbf{D} \vec{u}_j$, and $\gamma_j=\vec{u}_j^{\dagger} \mathbf{L} \vec{u}_j$, we get
\begin{equation}
    \alpha_j \Lambda_j^2 + \beta_j\Lambda_j + \gamma_j= 0,
    \label{eq:Lambda_pm}
\end{equation}
as in \cite{coletta2016linear}. Note that since $\bm{M}$ and $\bm{D}$ are diagonal, real and positive matrices, $\alpha_j, \beta_j>0$. Furthermore $\gamma_j$ must be real since $\bm{L}$ is a symmetric real matrix. 

In principle, since $j=1,\dots,2N$, there are $4N$ solutions for $\Lambda$ fulfilling (\ref{eq:Lambda_pm}), twice the number of actual eigenvalues. In fact, this derivation from the eigenvector equation provides a necessary condition for the eigenvalues but not a sufficient one, namely, not all solutions to \eqref{eq:Lambda_pm} are eigenvalues. In the simple case where node damping is proportional to inertia, we can see that Eq. (\ref{eq:phase_eigenvectors}) for the phase components of the eigenvector associated to $\Lambda^+$ is identical to that for those associated to $\Lambda^-$. The eigenvectors differ in the frequency components. Therefore, there are only $N$ values for coefficients $\alpha_i$, $\beta_i$, and $\gamma_i$, and no spurious solutions of (\ref{eq:Lambda_pm}).

Now, we focus on the case in which consumer nodes are inertia-less. The Laplacian matrix $\bm{L}$ can be written as
\begin{equation}
\bm{L} = \begin{bmatrix}\bm{L}_\mathrm{GG} & \bm{L}_\mathrm{GL} \\
\bm{L}_\mathrm{LG} & \bm{L}_\mathrm{LL}
\end{bmatrix},
\end{equation}
where subindices $\mathrm{G}$ ($\mathrm{L}$) refer to the generator (load) variables. Similarly 
\begin{equation}
\bm{M}=\begin{bmatrix}\bm{M}_\mathrm{GG} & \bm{0}_\mathrm{GL} \\
\bm{0}_\mathrm{LG} & \bm{0}_\mathrm{LL}
\end{bmatrix}, \;
\bm{D}=\begin{bmatrix}\bm{D}_\mathrm{GG} & \bm{0}_\mathrm{GL} \\
\bm{0}_\mathrm{LG} & \bm{D}_\mathrm{LL}
\end{bmatrix}.
\end{equation}
The Jacobian is given by
\begin{equation}
\renewcommand\arraystretch{1.3}
    \bm{J}= \begin{bmatrix}\bm{0}_\mathrm{GG} & \bm{0}_\mathrm{GL} & \mathbb{I}_\mathrm{GG}\\
    -\bm{D}_\mathrm{LL}^{-1}\bm{L}_\mathrm{LG} & -\bm{D}_\mathrm{LL}^{-1}\bm{L}_\mathrm{LL} & \bm{0}_\mathrm{LG}\\
    -\bm{M}_\mathrm{GG}^{-1}\bm{L}_\mathrm{GG} & -\bm{M}_\mathrm{GG}^{-1}\bm{L}_\mathrm{GL} & -\bm{M}_\mathrm{GG}^{-1}\bm{D}_\mathrm{GG}\end{bmatrix}.
\end{equation}
The Jacobian eigenvectors fulfill
\begin{equation}
    \bm{J} \begin{bmatrix} \vec{u}_{\mathrm{G} j} \\ \vec{u}_{\mathrm{L} j} \\ \vec{w}_{\mathrm{G} j} \end{bmatrix} = \Lambda_j \begin{bmatrix} \vec{u}_{\mathrm{G} j} \\ \vec{u}_{\mathrm{L} j} \\ \vec{w}_{\mathrm{G} j} \end{bmatrix},
\end{equation}
where $\vec u_{\mathrm{G} j}$ and $\vec u_{\mathrm{L} j}$ are the generators' and consumers' phase components of the eigenvector $j$ ($j=1,\dots,N+N_G$), and $\vec{w}_{\mathrm{G} j}$ are the generators' frequency components. This implies
\begin{align}
     &\vec{w}_{\mathrm{G} j} = \Lambda_j {\vec u}_{\mathrm{G} j}, \label{eq:eigenvectors_phase_gen} \\
     & -\bm{L}_\mathrm{LG} \vec{u}_{\mathrm{G} j} -\bm{L}_\mathrm{LL} \vec{u}_{\mathrm{L} j} = \Lambda_j \bm{D}_\mathrm{LL} \vec{u}_{\mathrm{L} j} 
\label{eq:eigenvectors_phase_load}\\
     &-\bm{L}_\mathrm{GG} \vec{u}_{\mathrm{G} j}  - \bm{L}_\mathrm{GL}\vec{u}_{\mathrm{L} j} =\Lambda_j \bm{D}_\mathrm{GG}\vec{u}_{\mathrm{G} j} + \Lambda_j^2 \bm{M}_\mathrm{GG} \vec{u}_{\mathrm{G} j}.
     \label{eq:eigenvectors_freq_gen}
\end{align} 
Eqs.~(\ref{eq:eigenvectors_phase_load}) and (\ref{eq:eigenvectors_freq_gen}) can be rewritten as
\begin{equation}
-\bm{L}\vec{u_j} = \Lambda_j \bm{D} \vec{u_j} + \Lambda^2_j \bm {M} \vec{u_j},
\end{equation}
which reproduces Eq.~(\ref{eq:phase_eigenvectors}). Therefore in case of inertia-less consumer nodes, eigenvalues still fulfill (\ref{eq:Lambda_pm}). The difference is that now, since the rank of $\bm{M}$ is $N_G$, there are $N-N_G$ eigenvalues for which $\alpha_j=0$. Thus they are real and given by
\begin{equation}
    \Lambda_j=-\frac{\gamma_j}{\beta_j}.
    \label{eq:singlet_eigenvalues}
\end{equation}
Note that these eigenvalues depend mostly on the damping and much less on the inertia, only through the shape of the eigenvector.

The dynamical system (\ref{eq:oscillator}) always has one zero eigenvalue due to the phase rotational symmetry: adding the same constant value to all phases does not change the dynamics nor the state.
The phase components of the eigenvector associated to this mode are given by $\vec{u}_1 =  [1, 1, \dots, 1]^{T}/\sqrt{N}$, while the frequency components are $\vec{w}_1=\vec 0$. Since $\bm{M}^{-1}\bm{L}$ is a zero row sum matrix, $\vec{u}_1$ is an eigenvector with $\lambda_1=0$. 

For the simple case where Eq.~(\ref{eq:damping_propto_inertia}) is fulfilled, Eq. (\ref{eq:eigenvalues_proportionality_case}) for $\lambda_1=0$ leads to $\Lambda_1^+=0$ and $\Lambda_1^-=-\kappa$, which are the Jacobian neutral eigenvalue and the ratio of damping to inertia respectively. The eigenvector associated to $\Lambda^-$ is $(\vec{u},\vec{w})= 1/\sqrt{N(1+\kappa)}[1,...,1,-\kappa, ...,-\kappa]^T$. For the general case where (\ref{eq:damping_propto_inertia}) is not fulfilled, there will not be an eigenvalue equal to $-\kappa$. Conventional power plants typically have inertia and damping proportional to their nominal capacity, although the proportionality coefficient depends on the plant technology. Besides, low inertia or inertia-less consumer nodes typically have a rather small damping. Therefore, while actual grids do not satisfy Eq.~(\ref{eq:damping_propto_inertia}), eigenvalues and eigenvectors are still relatively close to those encountered in that homogeneous case.

From Eq.~(\ref{eq:Lambda_pm}) it follows that the real part of any pair of complex conjugate eigenvalues is $\mathrm{Re}(\Lambda_j^{\pm})= - \beta_{j} / 2\alpha_{j}$, thus it is always negative and, as a consequence, fixed points of Eq.~(\ref{eq:oscillator}) cannot undergo a Hopf bifurcation.
Instead, instabilities arise from a real eigenvalue crossing zero, generically associated to a saddle-node (SN) bifurcation in which stable fixed points are either created or annihilated. Whether the real eigenvalue crossing zero is part of a pair given by Eq.~(\ref{eq:Lambda_pm}) or is a singlet given by  Eq.~(\ref{eq:singlet_eigenvalues}), it requires $\gamma_j=0$ which corresponds to a second zero eigenvalue of $\bm{L}$.

As shown in Ref.~\cite{bronski2014spectral} (Lemma 2.8), for a connected weighted graph, a second eigenvalue of $\mathbf{L}$ vanishes when
\begin{equation}
    \sum_{T} \pi(T)=0, 
    \label{eq:sum_weighted_trees}
\end{equation}
where the sum runs over all spanning trees $T$ of the graph and 
\begin{equation}
    \pi(T)=\prod_{(i,j)\in T} L_{ij}.
     \label{eq:weighted_tree}
\end{equation}
Since trees do not contain self-loops, only the off-diagonal coefficients of the Laplacian contribute to Eq.~(\ref{eq:weighted_tree}). If the phase differences for all lines are smaller than $\pi/2$,
according to Eq.~(\ref{eq:Laplacian}), all $L_{ij, i\neq j}$ are negative. Then, since all spanning trees of a graph with $N$ nodes have $N-1$ edges, all products $\pi(T)$ have the same sign: positive for $N$ odd and negative for $N$ even; and the sum (\ref{eq:sum_weighted_trees}) can not be zero. However, as soon as a line $l_s$ reaches a  $\pi/2$ phase difference, $\pi(T)=0$ for all spanning trees including $l_s$. There are two scenarios. First, if $l_s$ is the unique line connecting a part of the grid with the rest (for instance a leaf node or a cul-de-sac configuration), it is part of all spanning trees and thus all $\pi(T)$ become zero simultaneously. There is grid separation into two areas, each with its own Laplacian and the corresponding zero mode, and the frequency in one area deviates from the one in the other area, destroying synchrony between the two areas. This is similar to when a graph gets disconnected, hence the name "algebraic connectivity" given to the first non-zero eigenvalue of the Laplacian \cite{fiedler1975property}. Second, if $l_s$ is not the unique line connecting a part of the grid to the rest, its phase difference can be larger than $\pi/2$. Then, $\pi(T)$ changes sign for all the spanning trees including $l_s$. At some point this can lead to the sum being zero, which destabilizes the fixed point. These two scenarios qualitatively explain how stability is lost when one or several phase differences exceed $\pi/2$. 

For the very unlikely case where two or more lines simultaneously reach a $\pi/2$ phase difference, the first scenario applies if all the lines connecting one sub-set of the grid with its complement sub-set are within the set of lines that have reached $\pi/2$. Otherwise, the second scenario applies.

Typically, the nominal capacity of a line is first restricted by its thermal limit, which is often much lower than the maximum flow $b_{ij}$. Transmission power grids are designed to operate normally at phase differences smaller than $\pi/4$. Therefore, regimes where phase differences are close to $\pi/2$ cannot be sustained for long periods of time. However, large phase differences can be reached during short periods. It is important 
to maintain the stability of the power grid under these conditions to avoid the loss of synchrony during a contingency.

\section{Imbalances in the European Grid} \label{sec:European_grid}

We apply the previous analysis to a large heterogeneous grid, where instabilities are generally not associated to line overloads. In particular, we consider PanTaGruEl \cite{pagnier2019inertia,tyloo2019key}, an Open-Access model \cite{pantagruel} for the synchronous grid of continental Europe. It considers a total of $N=3809$ buses, of which $N_{\rm G}=468$ are operative generators, and $4944$ power lines.
Inertia, damping, and line parameters are heterogeneous. The generation and load of each bus is obtained with an optimal power flow using publicly available data from ENTSO-E \cite{entsoedata}. Therefore, this is a realistic representation of the actual grid.

In the next sections, we analyze how increasingly large generation imbalances between different regions affect the stability of the grid. We focus on the power balance only, assuming there is enough control capacity. We explore various imbalance scenarios derived from different reference dispatches.

Each reference dispatch is obtained using the optimization algorithm described in \cite{pantagruel}. This algorithm calculates the generation and the load at each node from ENTSO-E load data at country level for a specific date and time. Therefore, each dispatch is characterized by a different load and generation configuration.

Then, to obtain the different generation configurations, we reduce the generation in a region R by a fraction $r$ and increase it in a region I by $i$ such that the total generation remains unchanged. Selecting these regions may be done manually by choosing key areas in the network, or in a systematic way. The results presented here follow the first approach, while the latter is used in the Supplemental Material to support the generality of our method.

The reference dispatch is a fixed point of the system. If we modify the generation distribution, the power flow will also change and with it the voltage phases at each node. Thus, we have to calculate the new fixed points of the system for each new generation configuration. Using Newton's method, we calculate and continue the fixed point for increasing values of $r$. Then, we analyze the linear stability of the fixed point by computing the eigenvalues and eigenvectors of the Jacobian matrix evaluated at that point.

\begin{figure}[H]
    \centering
    \includegraphics[width=\linewidth]{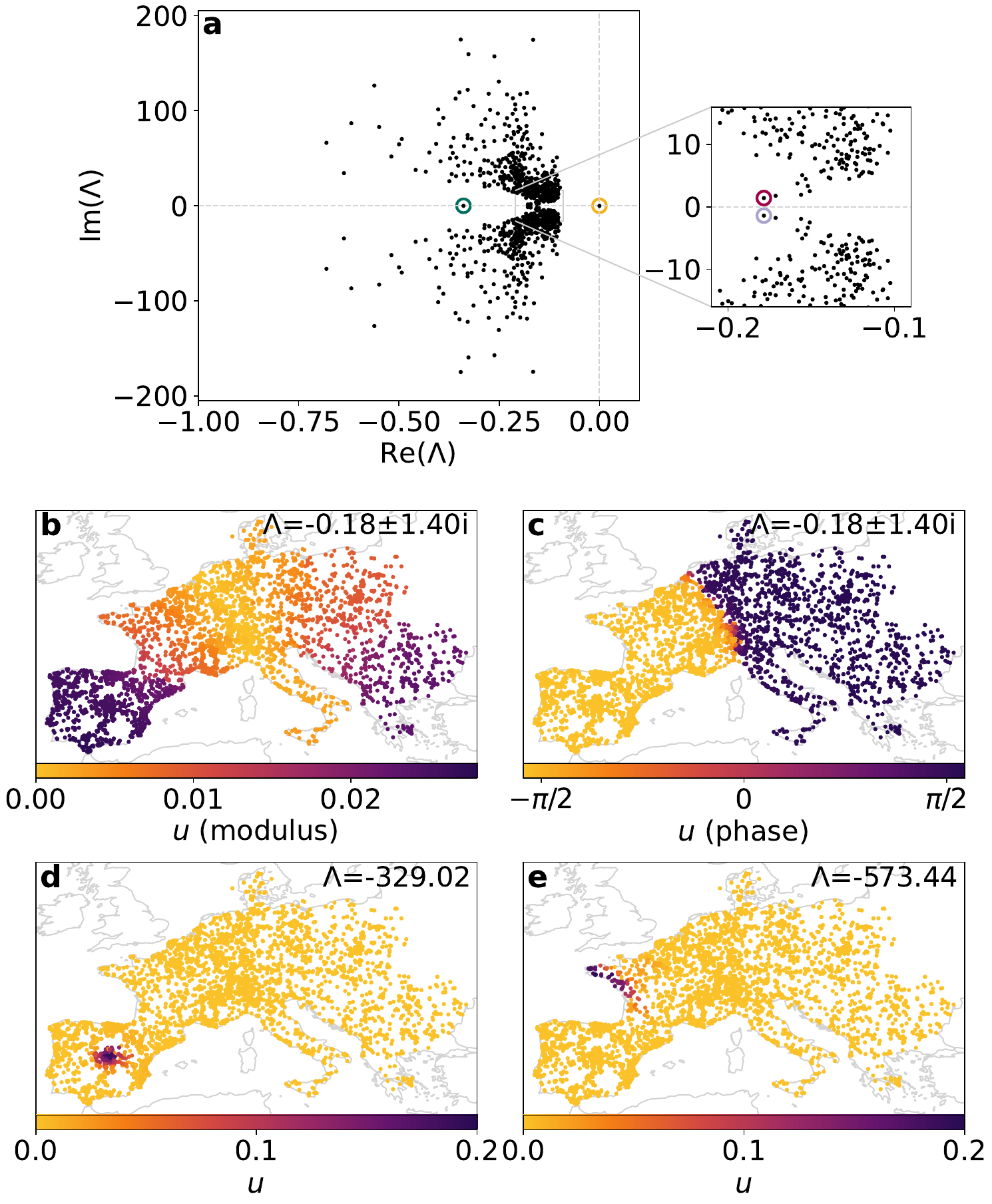}
    \caption{(a) Least damped eigenvalues of the Jacobian for the reference case. Phase components of some eigenvectors: (b,c) lowest frequency, (d) second and (e) third least damped real modes. }
    \label{fig:eigval_eigmodes}
\end{figure}

We briefly discuss the Jacobian eigenvalues and eigenvectors corresponding to the chosen reference dispatch. 
FIG.~\ref{fig:eigval_eigmodes}(a) shows the least damped eigenvalues of the Jacobian, obtained from a linear stability analysis of the reference case. The real eigenvalue arising from the neutral phase eigenvector of the Jacobian, $\Lambda=0$, and the one with an eigenvalue close to $\Lambda=-\sum d_i/\sum m_i=-0.339$ are plotted in yellow and green respectively. There are an additional $3351$ real eigenvalues which come from phase eigenvectors with tiny components in the generator nodes, $\vec{u}_{\mathrm{G} j} \approx 0$, given by Eq.~(\ref{eq:singlet_eigenvalues}). They are more strongly damped, with a real part that is more negative, outside of the range of the horizontal axis of FIG.~\ref{fig:eigval_eigmodes}(a). The remaining $2(N_{\rm G}-1)=914$ eigenvalues are complex and have relatively small damping, given by $-\beta_j/2\alpha_j$, and they are all shown in FIG.~\ref{fig:eigval_eigmodes}(a).

In FIG. \ref{fig:eigval_eigmodes}, we further take a closer look at some relevant modes. All eigenvectors have a phase component for each bus and a frequency component for each generator. On the one hand, panels (b) and (c) show the phase components of the eigenvectors with smallest imaginary part of their eigenvalues, i.e. either the magenta or lilac eigenvalue in FIG.~\ref{fig:eigval_eigmodes}(a). We plot modulus and phase because this is a complex mode, hence it has imaginary components. We observe that this mode has a large constant amplitude in the Iberian Peninsula and the Balkans, while it is zero in Central Europe. Thus, when it is excited, it creates large antiphase coherent oscillations between east and west. These oscillations are known in the literature as inter-area oscillations \cite{fritzsch2022long}. This mode is associated with the slowest mode of the Laplacian. Quite interestingly, it is however not the least damped mode of the Jacobian due to the heterogeneous distribution of inertia and damping across the network. Although not shown here, the frequency components are similar to the phase components.

On the other hand, panels (d) and (e) show the phase components of two of the least damped real modes, lying outside of the horizontal range of FIG.~\ref{fig:eigval_eigmodes}(a). Since they are real, the eigenvectors only have real components. These two modes in particular are localized in Madrid and in the French Atlantic Coast, which are areas with several consumer buses but very few generators. In fact, the projection of the inertia matrix on these modes is very small. Their frequency component is also very small throughout the network except in the few generators close to these locations, which determine the response to frequency fluctuations. Although we only show these two modes in particular, the discussion is general for all real modes.

\section{North-South and Central-East imbalances}
\label{sec:north-south_central-east}

\begin{figure}[H]
    \centering
    \includegraphics[width=.7\linewidth]{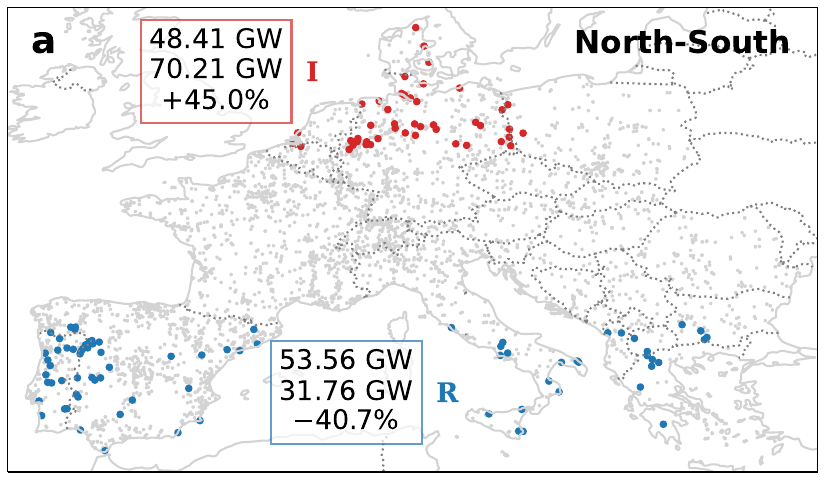}
    \includegraphics[width=.7\linewidth]{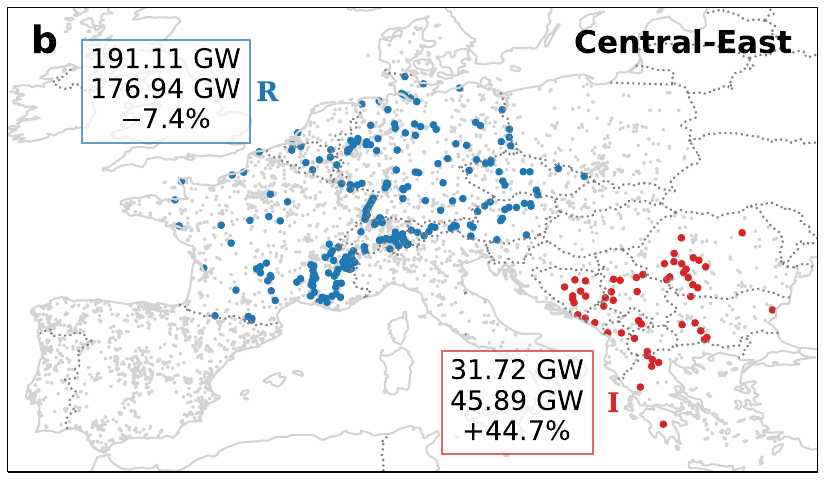}
    \caption{Power imbalance regions in the (a) North-South and (b) Central-East cases. Blue and red dots mark generators where power is reduced (R) and increased (I) respectively. Each text box displays the initial, final, and percentage change in generation of the marked nodes.}
    \label{fig:northsouth_centraleast1}
\end{figure}

We analyze different imbalance scenarios but focus on the two cases shown in FIG. \ref{fig:northsouth_centraleast1}. On the one hand, motivated by the significant growth in wind power generation, both onshore and offshore, in the North of Europe, we explore scenarios where generation increases in the northern regions and decreases below a certain latitude, and vice versa. Countries such as Denmark, Germany, and the Netherlands, have favorable geographical conditions for strong and consistent winds, which make these areas prime locations for wind farms. Conversely, the South of Europe has ideal conditions for photovoltaic generation. Therefore, in a more renewable grid, larger power flows are expected between these two areas, depending on meteorological conditions \cite{hofmann2018principal}.

\begin{figure}[H]
\centering
\includegraphics[width=.9\linewidth]{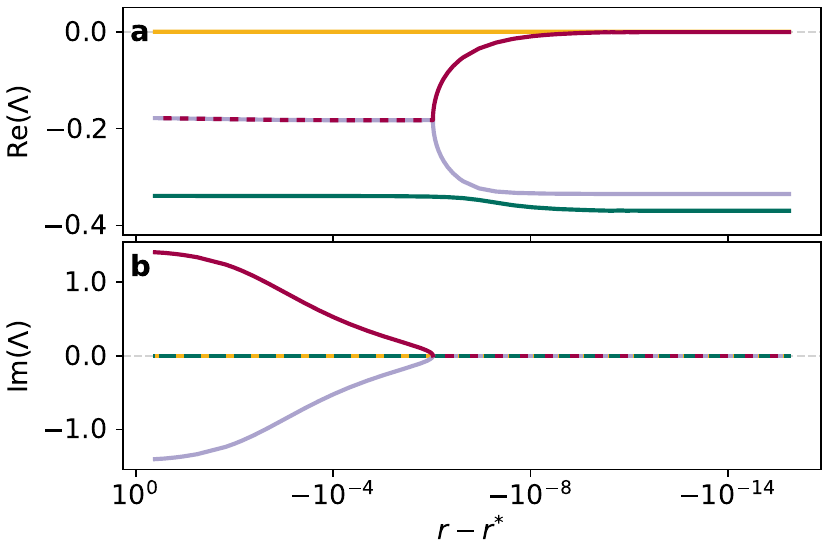}
\caption{Evolution of (a) the real and (b) the imaginary parts of the zero eigenvalue (circled in yellow), least damped real one (green), and the complex conjugate pair (lilac and magenta) involved in the BD transition and triggering the SN bifurcation. The horizontal axis is plotted in logarithmic scale with respect to the value $r^*$ at which the SN bifurcation takes place and the stable, synchronous fixed point disappears. }
\label{fig:eigval_evolution}
\end{figure}

On the other hand, in 2021, several lines in the Balkans overloaded, splitting
the grid in two disconnected subsets~\cite{entsoe2021}. This situation was caused by a lower demand in the South-East and a higher demand in Central and Western Europe due to weather conditions. To replicate the 2021 situation, we reduce the generation in Central Europe, while simultaneously increasing it in the Balkan region, as shown in FIG. \ref{fig:northsouth_centraleast1}(b). 
Later, also in 2021, the Iberian Peninsula separated from the rest of the continent, following a cascade of line trips initially triggered by a forest fire~\cite{entsoe2022}.


As the power imbalance increases, the system loses stability through a SN bifurcation at $r^*=0.40703$ (North-South imbalance case) and $r^*=0.07416$ (Central-East imbalance case). While this bifurcation involves a real eigenvalue reaching zero, it does not come from any of the real eigenvalues given by Eq.~(\ref{eq:singlet_eigenvalues}) with small components on nodes with inertia. Rather, the scenario is as shown in FIG. \ref{fig:eigval_evolution} for the North-South case. The situation is similar for all other cases we have investigated. Purely real eigenvalues are too damped to lead to instabilities. Instead, a pair of 
weakly damped, complex conjugated eigenvalues see their imaginary part decrease as $r$ increases, until it becomes zero and the eigenvalues become real. This is a Belyakov-Devaney (BD) transition \cite{devaney1976reversible,homburg2010homoclinic}. As $r$ increases further, 
the two resulting real eigenvalues repel each other, thus one of them 
moves towards positive values. This causes a SN bifurcation when the eigenvalue reaches zero. Since eigenvalues separate very fast with $r$ after the BD transition, the SN bifurcation is located very close to the BD transition.


\begin{figure}[t]
\centering
\includegraphics[width=\linewidth]{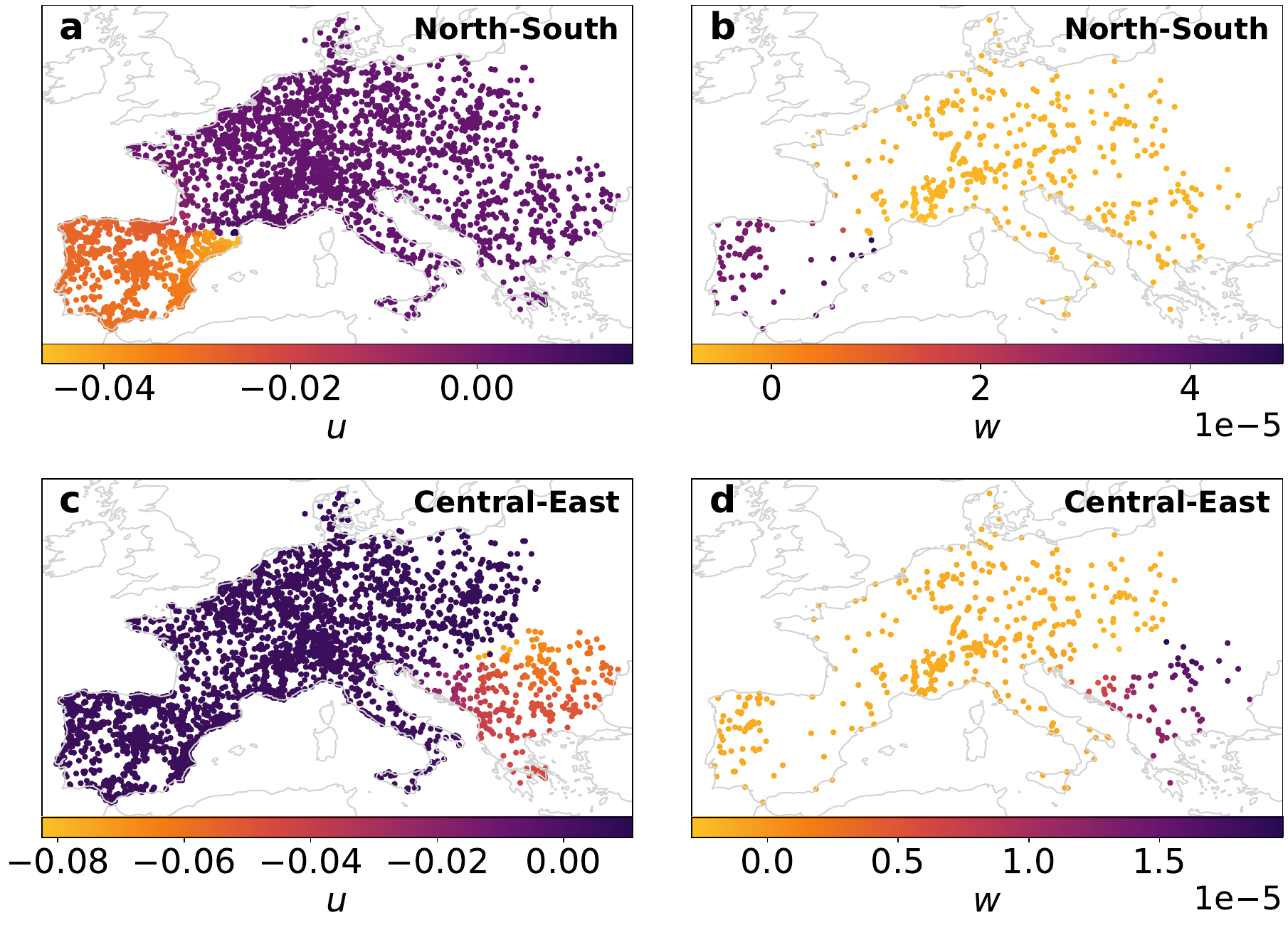}
\caption{(a,c) Phase and (b,d) frequency components of the eigenvector associated with the eigenvalue that reaches zero at the SN. (a,b) North-South and (c,d) Central-East imbalance cases.}
\label{fig:northsouth_centraleast_eigvec}
\end{figure}

In FIG. \ref{fig:northsouth_centraleast_eigvec}, we show the eigenvectors associated with the eigenvalues that reach zero in the North-South [Panels (a) and (b)] and Central-East cases [Panels (c) and (d)]. Note that, initially, these modes are similar to the one in FIG. \ref{fig:eigval_eigmodes}(b) and \ref{fig:eigval_eigmodes}(c). As the interregional power imbalance increases, the mode changes, showing that the grid splits into two areas when the system reaches instability. 
We observe that the splitting does not necessarily occur precisely around the region where the power increases. The power is carried from that area to elsewhere, causing large power flows throughout the grid.

On the one hand, in the North-South case, Spain and Portugal have limited connectivity to France and the rest of the grid given their geographical location. This area is known to be a bottleneck of the European grid. In fact, the Iberian Peninsula has been historically considered an electric island \cite{entsoeCSW}. On the other, the Balkan region is vulnerable in particular due to historical reasons, most notably the electric infrastructure inherited from the Socialist Federal Republic Yugoslavia \cite{balkans}.


\begin{figure}[t]
\centering
\includegraphics[width=.9\linewidth]{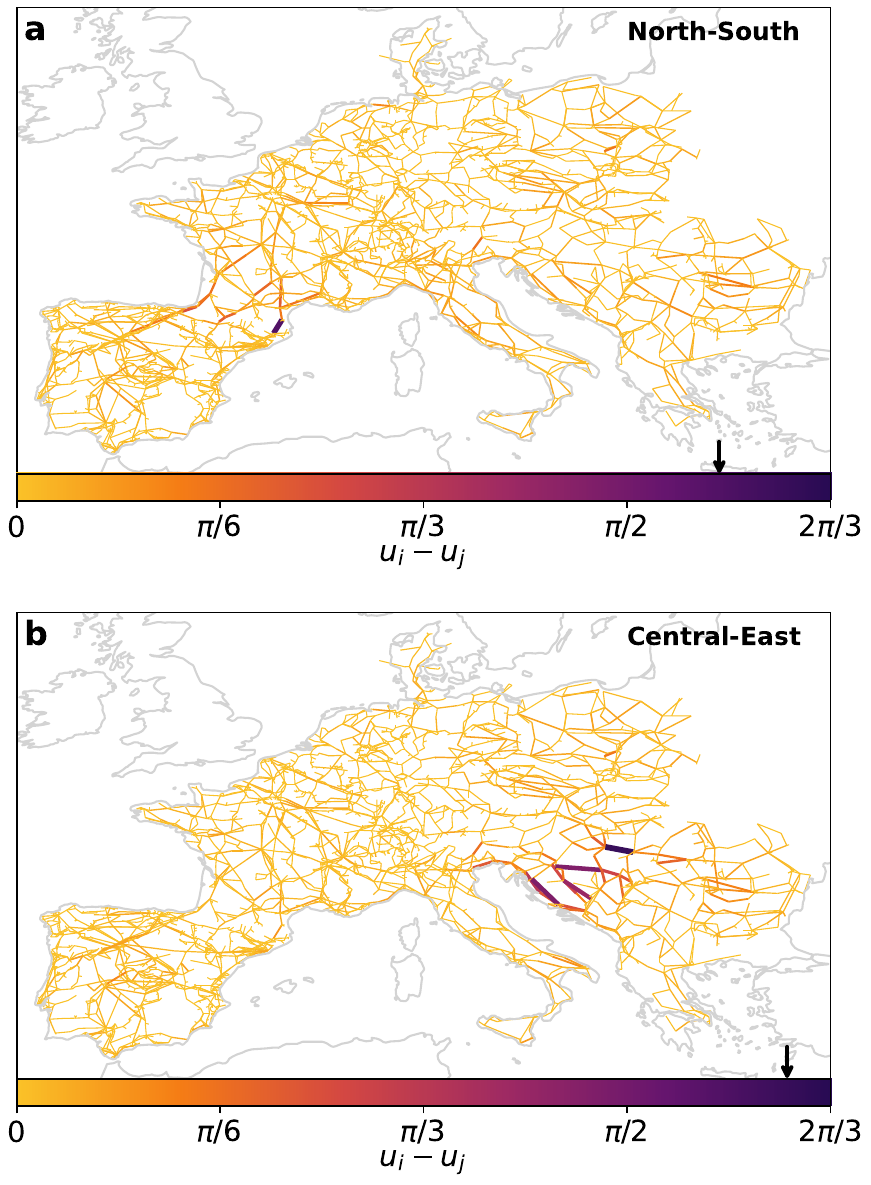}
\caption{Line phase differences at the threshold of instability. The arrow in the color scale signals the maximum phase difference. (a) North-South and (b) Central-East imbalance cases.}
\label{fig:northsouth_centraleast_fixedpoint}
\end{figure}

In FIG. \ref{fig:northsouth_centraleast_fixedpoint}, we show the phase differences of the lines just before the SN bifurcation for the North-South and Central-East cases. Initially, all lines are below $\pi/6$. However, before the bifurcation, we observe that the phase difference of one line is above $\pi/2$ in both cases. This was also previously observed by Manik et al. in \cite{manik2014supply} for homogeneous inertia. For a $\pi/2$ phase difference, the line flow is maximum. Beyond $\pi/2$, the flow carried by the line decreases as the phase difference increases. Hence, the other lines have to carry more flow to compensate. The instability takes place when the flow carried by the lines is not enough to fulfill the needs of the more highly loaded area. It is not due to a capacity saturation, but to a restriction on the phase differences. Thus, the problem does not lie in having insufficient transmission capacity, but rather in the topology of the grid itself.

In the North-South imbalance case, the power imbalance in the Iberian Peninsula is negative because, there, the demand is higher than the generation. Thus, the Peninsula needs to import power through the four lines that connect it to France. The instability occurs when the imbalance in the Peninsula is increased from $-2.92$ GW (reference) to $-14.92$ GW. At that point, the line \textit{Baixas 380 - Vic 380} [line \#1 in FIG. \ref{fig:northsouth_centraleast_fixedpoint_DC}(b), to be discussed in Section \ref{sec:DC_trick}] has a phase difference of $103.6^{\circ}$, i.e. it is at $97.2\%$ of its capacity. For the other three lines, the phase difference is below $44.1^{\circ}$ and the flow is less than $70\%$ of the maximum. The power flow of $14.92$ GW that these four lines carry to balance demand and generation in the Peninsula, remains well below their total transmission capacity of $20.74$ GW.

In the Central-East imbalance case, the line \textit{Paks 380 - Sandorfalva 380} [line \#1 in FIG. \ref{fig:northsouth_centraleast_fixedpoint_DC}(d)] has a phase difference above $90^{\circ}$ at the threshold of the instability. The next largest phase differences are observed in lines \textit{Brinje 220 - Konjsko 220} (\#2) and \textit{Ernestinovo 380 - Unlabeled\_290 380} (\#3), which are at $85.5^{\circ}$ and $84.0^{\circ}$ respectively. In terms of power flow, several lines are close to their limits: \textit{Brinje 220 - Konjsko 220} is at $99.7\%$, \textit{Ernestinovo 380 - Unlabeled\_290 380} at $99.5\%$, \textit{Prijedor 220 - Kakanj 220a} (\#4) at $96.3\%$ and \textit{Paks 380 - Sandorfalva 380} at $91.6\%$. 
Quite interestingly, our numerics reproduce the same grid splitting 
as the 2021 grid separation into two areas for both the Central-East~\cite{entsoe2021} and the North-South imbalance~\cite{entsoe2022}. Furthermore, for the Central-East imbalance case,
some of the substations involved in the actual incident, e.g. \textit{Brinje}, \textit{Ernestinovo}, and \textit{Prijedor}, also stand out in our analysis for having large power flows.


\begin{figure}[t]
    \centering
    \includegraphics[width=\linewidth]{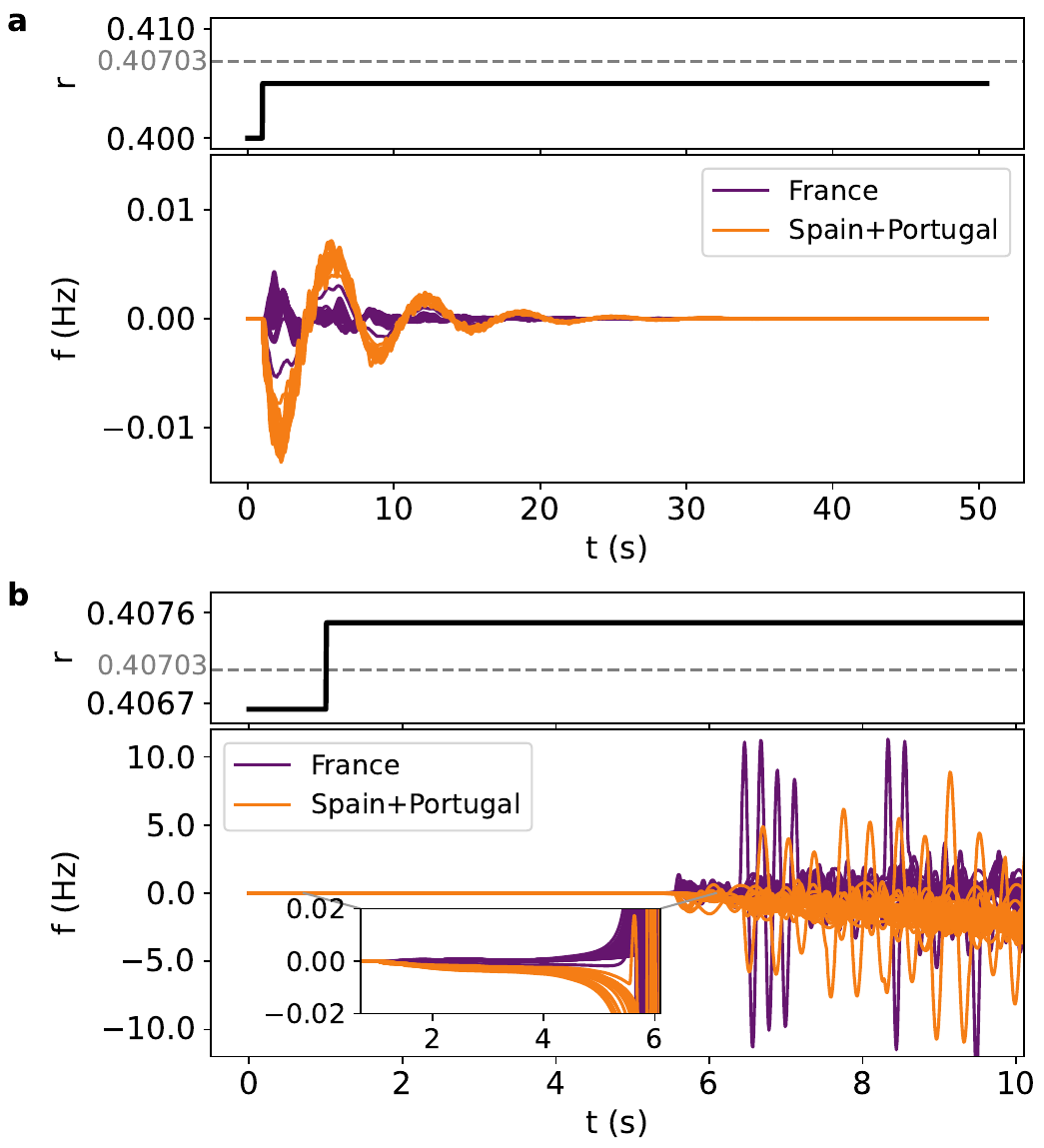}
    \caption{Frequency of generators in Spain and Portugal (orange lines) compared to those in France (purple lines) after increasing, at time $t=1$ s, the imbalance ratio $r$ to a value (a) below or (b) above $r^{*}$ in the North-South case.}
    \label{fig:northsouth_freq}
\end{figure}

To further illustrate the emergence of instability, we finally run dynamical simulations for the North-South imbalance, starting from a fixed point at a value $r<r^*$ of the power imbalance below the critical value, 
giving an instantaneous increase to a larger value of $r$ and letting the system evolve. The results are shown in FIG. \ref{fig:northsouth_freq}.  In panel (a), the final value of $r$ is still below $r^*$, hence the system reaches a new stable fixed point, following  a short transient. However, in panel (b), where the final value of  $r$ is above $r^*$, the system does not reach a new fixed point. Instead, the frequency first decreases 
in the Iberian peninsula and increases in France. As shown in the inset, this causes the system to split into two unbalanced asynchronous areas. Finally, the frequency reaches an incoherent state with very large, undamped frequency oscillations.

\section{Forcing grid splitting: the DC trick} \label{sec:DC_trick}

The instability is dictated by topology constraints arising from the fact that the power transmitted by AC lines depends on the phase difference between the connected buses. 
In contrast, the power transmitted by DC lines does not suffer from such constraint. 
With this in mind, we propose a method to stabilize the system beyond the instability observed with AC lines, which eventually forces the system to split into disconnected areas. In this way, we can identify sets of critical lines potentially leading to such splitting when circuit breakers and other safety control devices are taken into account.

The method works as follows. In AC power grids, the voltage on each node has a well defined phase $\theta_{i}$. Pairs of nodes connected by a power line thus define a phase difference 
$\theta_{i} -\theta_{j}$. For each line in the grid, when 
this phase differences is smaller that $\pi/2$, the line operates as a regular AC line, and the power flow it carries is proportional to $\sin (\theta_{i} -\theta_{j})$. When loads or interregional flows increase, phase differences increase until one or several of them reach 
and exceed $\pi/2$. When this occurs, these lines are artificially turned into DC lines, 
with a power flow remaining at its maximal value, regardless of 
the phase of the connecting buses. Formally, in the dynamical system defined in Eq.~(\ref{eq:oscillator}), we replace the flow terms $b_{ij} \sin(\theta_i-\theta_j)$ by $F_{ij}$ defined by
\begin{equation}
    F_{ij}= \left\{
    \begin{array}{cc}
        b_{ij} \sin(\theta_i-\theta_j) & \text{for } |\theta_i-\theta_j|  \le \pi/2 \\ b_{ij} &  \text{for } |\theta_i-\theta_j|  > \pi/2
    \end{array} \right. .
\end{equation}
In principle, 
such mathematical trick could be implemented using High-Voltage DC (HVDC) lines, where the power electronics of converters can be programmed to deliver any desired power. 
Obviously, this scheme would
be impractical in real power grids and we stress that it is only meant to be an identification method. 
We call HVDC lines those with $|\theta_i-\theta_j|  > \pi/2$ in the upcoming investigations. 

It is important to realize that, from the point of view of the weighted network Laplacian of 
Eq.~\eqref{eq:Laplacian}, our method removes each line with $|\theta_i-\theta_j|  > \pi/2$.
Below we will see that, as interregional power imbalances increase, more and more lines 
are turned into DC lines, which eventually splits the grid defined by the weighted 
network Laplacian into disconnected areas. 
In this way, the method allows to identify a cutset, i.e. a set of lines that separates the grid into disconnected subgrids \cite{song2017network,song2017characterization}. The method extends the stability of the fixed point, and the lines that reach a $\pi/2$ phase difference one by one eventually reveal a natural segmentation of the grid.

Taking into account that, for $i\ne j$, the Laplacian is now given by
\begin{equation}
    L_{ij} = \left\{
    \begin{array}{cc}
        -b_{ij} \cos(\theta_i-\theta_j) & \text{for } |\theta_i-\theta_j|  \le \pi/2 \\ 0 &  \text{for } |\theta_i-\theta_j|  > \pi/2
    \end{array} \right. ,
\end{equation}
the stability analysis discussed in Section \ref{sec:model} holds. In particular, the instability takes place when Eq.~(\ref{eq:sum_weighted_trees}) is fulfilled. The difference is that now $L_{ij, i\neq j} \le 0$, thus $\pi(T)$ can not change sign and instabilities can only take place via the first scenario discussed above. As a consequence, applying this method, the fixed point is more robust and can be continued for larger generation imbalances.

\begin{figure}[H]
\centering
\includegraphics[width=\linewidth]{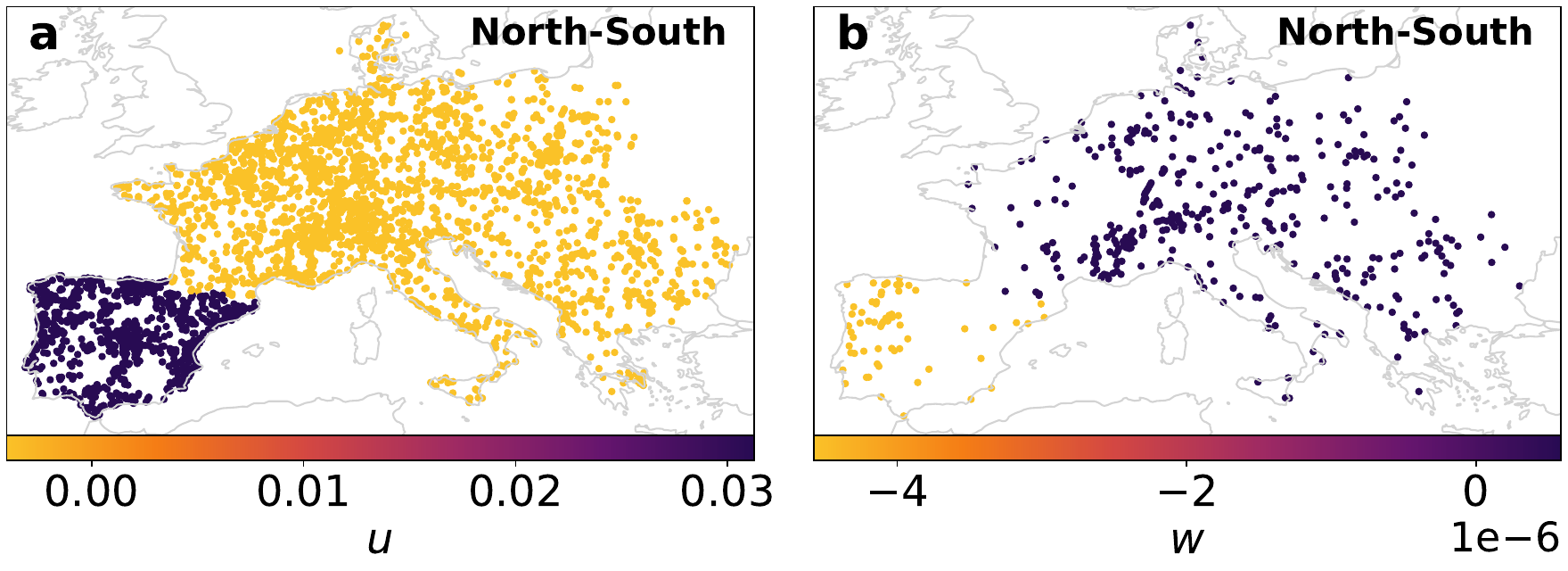}
\includegraphics[width=\linewidth]{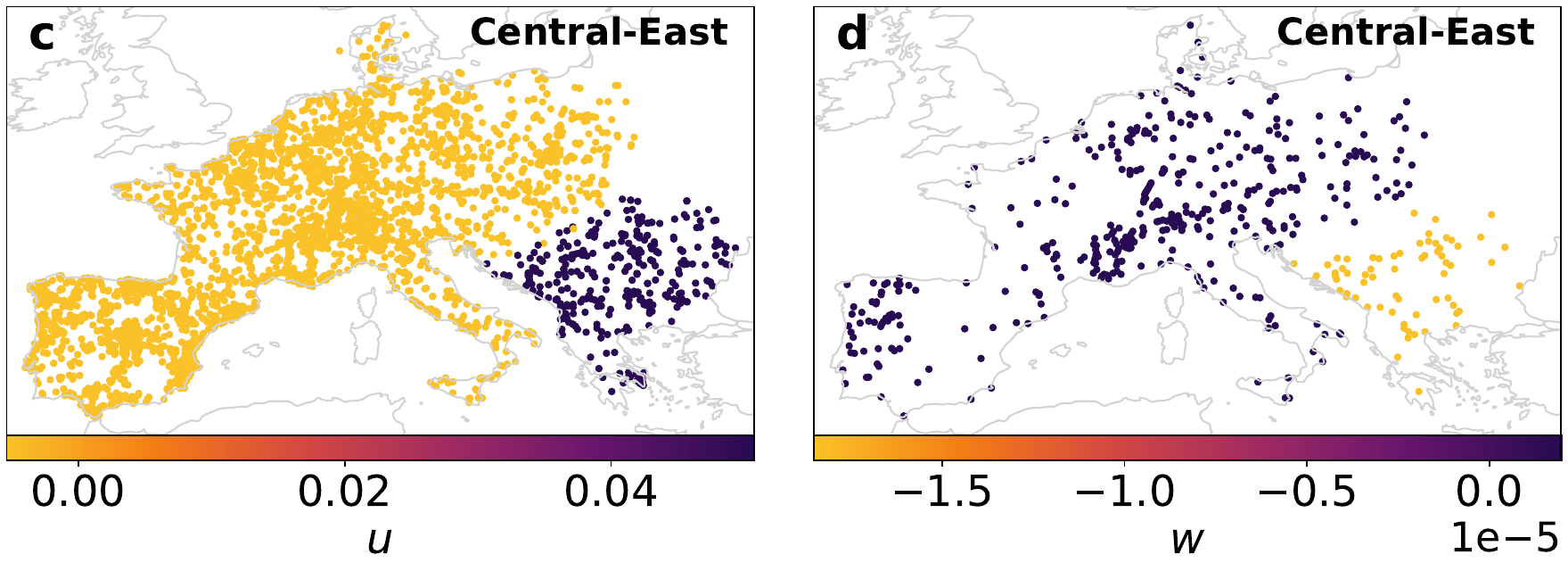}
\caption{(a,c) Phase and (b,c) frequency components of the eigenvector associated with the eigenvalue that becomes zero. (a,b) North-South and (c,d) Central-East cases.}
\label{fig:northsouth_centraleast_eigvec_DC}
\end{figure}

\begin{figure*}[t]
\centering
\includegraphics[width=.48\linewidth]{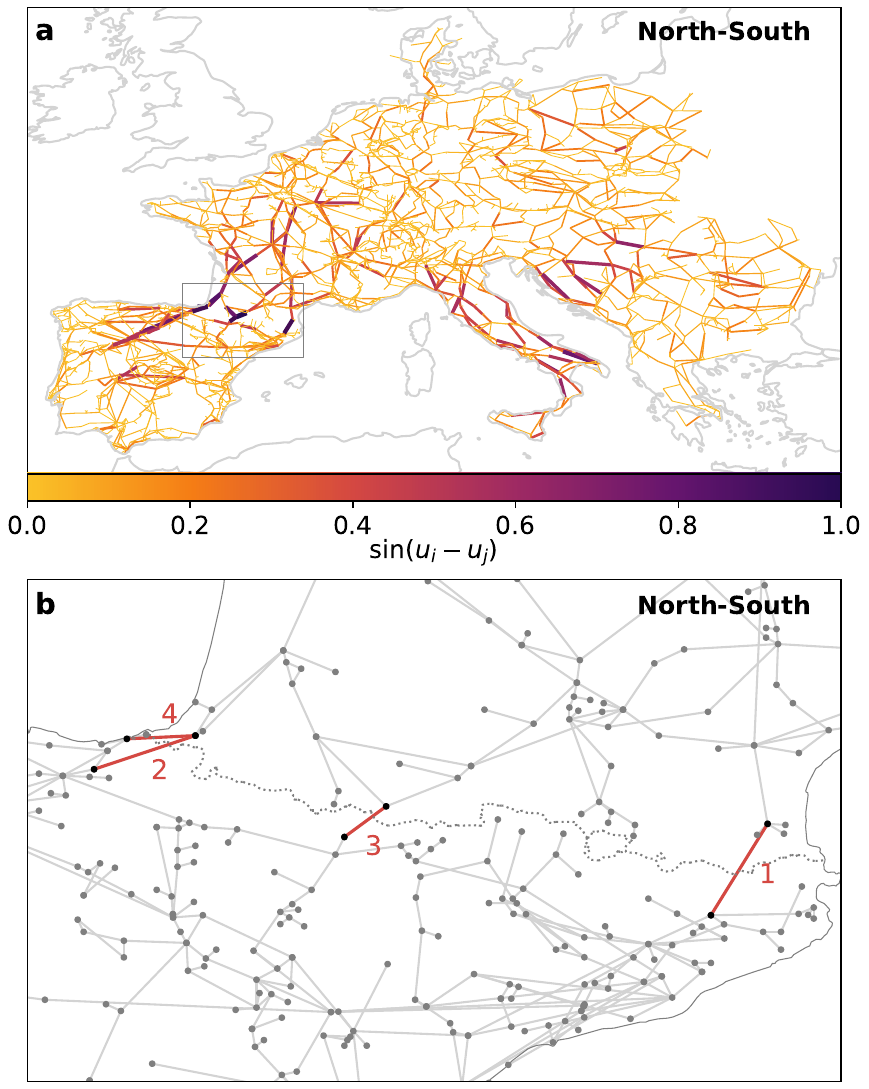}
\includegraphics[width=.48\linewidth]{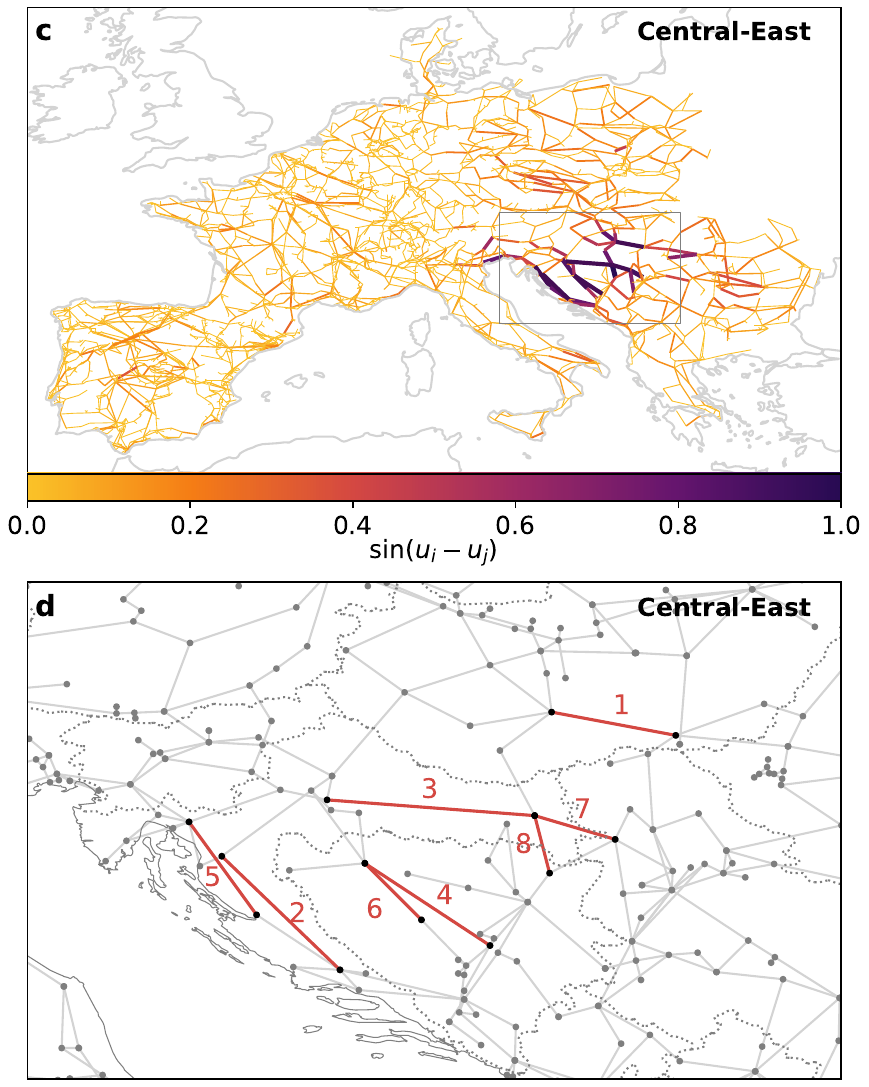}
\caption{(a,c) Line stress at the threshold of the instability. (b,d) Lines changed to DC in the splitting region labeled by order of conversion. (a,b) North-South and (c,d) Central-East cases.}
\label{fig:northsouth_centraleast_fixedpoint_DC}
\end{figure*}

Using this method, we can continue the fixed point from a critical imbalance of  $r^*=0.40703$ to $r^*=0.88772$ for the North-South case, and from $r^*=0.07416$ to $r^*=0.10040$ for the Central-East case. In panels (a) and (c) of FIG.~\ref{fig:northsouth_centraleast_fixedpoint_DC}, we show the stress of lines just before the instability, i.e. just before the lines separating two regions about to be disconnected
[the Iberian Peninsula from the rest of the continent in FIG.~\ref{fig:northsouth_centraleast_fixedpoint_DC}(a); the Balkans from the Center-West part of the 
Continent in  FIG.~\ref{fig:northsouth_centraleast_fixedpoint_DC}(c)]
reach a $\pi/2$ phase difference -- they reach their maximum capacity of $20.74$ GW for the Pyrenees and $27.72$ GW for the Balkans. When that happens, the network separates into two components and the Jacobian matrix acquires a new zero eigenvalue, whose eigenvector is shown in FIG.~\ref{fig:northsouth_centraleast_eigvec_DC}.

Finally, in panels (b) and (d) of FIG.~\ref{fig:northsouth_centraleast_fixedpoint_DC}, we zoom into the splitting regions, where HVDC lines are displayed in red and labeled according to the order in which they reach the critical phase difference of $\pi/2$. We can see that the Iberian Peninsula and the Balkans are disconnected from the rest of the grid, once those red links are removed. Hence, they become desynchronized by changing those lines from AC to controlled HVDC. It is worth noting that these lines are not the minimum set of lines needed to disconnect the two regions. For instance, keeping 
lines \# 4 and 6 in FIG.~\ref{fig:northsouth_centraleast_fixedpoint_DC} and removing a single line to the
west of them also leads to grid separation. 
Also, although the order of conversion may vary and additional lines may also be converted to DC, all splitting cases in these regions involve these critical lines.

\section{Final discussion and Conclusion} \label{sec:conclusions}

\begin{figure*}[t]
\centering
\includegraphics[width=.48\linewidth]{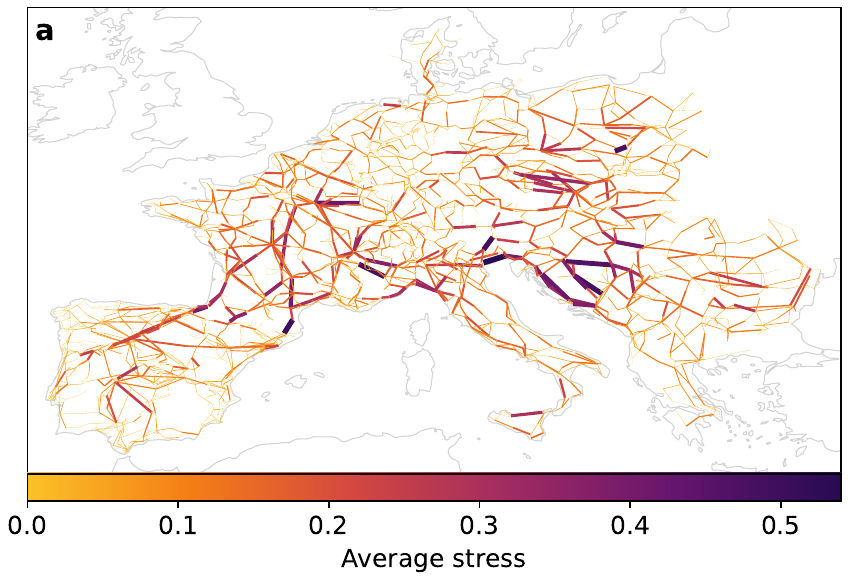}
\includegraphics[width=.47\linewidth]{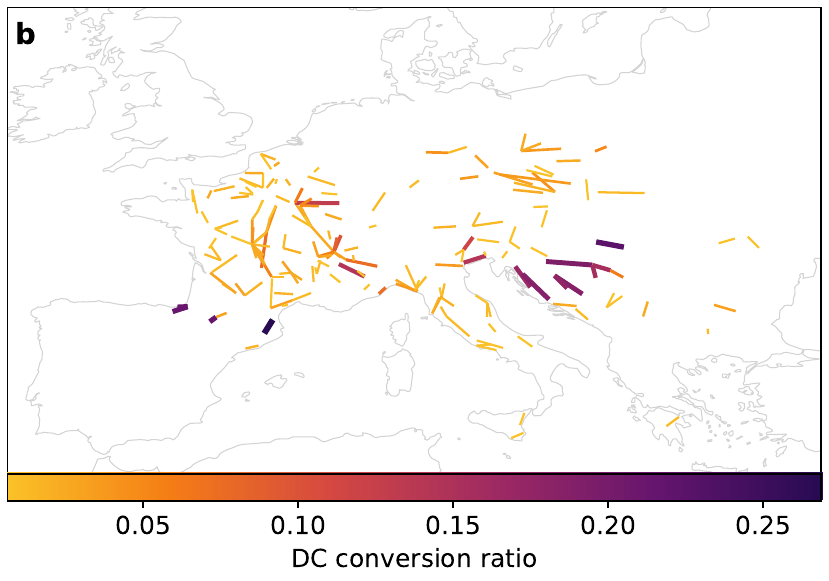}
\caption{(a) Line stress averaged over the 160 total number of cases. (b) Ratio between the number of cases in which a line is converted to DC over the total number of cases. See Supplemental Material for details on the analyzed scenarios.}
\label{fig:linestress_DCconversion}
\end{figure*}

Using a model of the synchronous power grid of continental Europe, we investigated various scenarios characterized by increasingly large interregional power flows. These could be caused by the large deployment of renewable energies in some regions in specific meteorological conditions. Beyond critical interregional flows, the system loses stability and our goal has been to identify critical lines that get overloaded first. We have developed a methodology for identifying said lines and observe that forced grid splitting generally does not occur at the boundary between the region where generation is increased and the region where it is decreased. Most notably, the splitting boundaries match in two instances those observed in real grid separation events in 2021~\cite{entsoe2021,entsoe2022}.

Beyond the two cases we discussed in detail above,  we have performed an extensive, systematic numerical analysis of power imbalances, covering a total of 160 scenarios, comprising 8 different dispatches, each paired with 20 different generation imbalances.
For a broad range of perturbations, splitting takes place either at the Pyrenees or the Balkans. In the Supplemental Material, we show the imbalance distributions that lead to each splitting. The instability mechanism is the same as explained in Sections \ref{sec:north-south_central-east} and \ref{sec:DC_trick}. Besides, we observe the same unstable eigenvectors, since their dot product is close to 1.

Similar to the case of the Iberian Peninsula, there are other locations where the geography shapes the topology of the grid, making it susceptible to disruptions for similar, obvious reasons. One such instance is the Italian Peninsula. However, we also observed separations in France and Central Europe where the grid is densely meshed. These results are included in the Supplemental Material. 

The method is therefore general and successfully identifies sets of critical lines
that are located at different places, depending on the considered dispatch and direction of imbalance. 
This is illustrated in FIG.~\ref{fig:linestress_DCconversion}.
In panel (a), we show the line stress at the last fixed point before the instability, which we calculate as an average over the 160 analyzed scenarios mentioned earlier. We complement this with the fraction of cases in which lines are converted to DC, displayed in panel (b). As expected from our previous discussion, we can observe that the Pyrenees and the Balkans are the two most vulnerable areas of the system. However, there are also other critical lines, for instance in France, limiting interregional power transmission.

Generally, we have found that instabilities occur well after one or several phase differences
exceed $\pi/2$. This means that line transmission capacity limits are not the immediate cause of the instability. Instead, the instability emerges from grid topology constraints. Specifically, it arises from the fact that power transmission through AC lines depends on the phase difference between connecting buses, and sometimes these cannot accommodate the necessary transmission power. Mathematically, the mechanism behind the instability is a pair of complex eigenvalues, associated to a Jacobian eigenvector with non-zero components on nodes with inertia. They become real (Belayakov-Devaney transition) and move apart, triggering a saddle-node bifurcation when one of them becomes zero.

To push the system beyond this limit, we have proposed a mathematical trick forcing the splitting of the
grid into disconnected areas. Our strategy involves converting certain lines to DC, which allows us to continue the fixed point further. By doing so, we are able to identify crucial lines in the splitting boundary. These lines are not necessarily the minimum set of lines connecting two given regions. For instance, in the Pyrenees, there are four critical lines and they are the minimum cutset. However, in the Balkans, we identify eight critical lines, while the minimum cutset would consist of seven.

Finally, we stress that our method -- which we applied to a model of the synchronous grid of continental Europe -- is general and can also be applied to identify the most critical lines in other large-scale transmission grids.

\begin{acknowledgments}
    M. Mart\'{i}nez-Barbeito, D. Gomila, and P. Colet acknowledge funding from the European Comission integrated action Virtual Power Plants for Islands, VPP4ISLANDS, (Grant agreement N\textsuperscript{o} 957852), from project APASOS PID2021-122256NB-C22 financed by MCIN/AEI /10.13039/501100011033/ and by EU through FEDER funds, and from Maria de Maeztu program CEX2021-001164-M of MCIN/AEI /10.13039/501100011033/.
    
    J. Fritzsch and P. Jacquod acknowledge funding from the Swiss National Science Foundation under grant 200020\_182050.
\end{acknowledgments}

%

\clearpage

    \input{supp}

\end{document}

%% file: supp.tex
\onecolumngrid

\section{Supplemental Material}
\beginsupplement

\subsection{Systematic analysis of interregional power imbalances}

We systematically analyze a total of 160 scenarios, comprising 8 different dispatches, each paired with 20 unique generation imbalances.

\begin{figure}[H]
    \centering
    \includegraphics[width=0.5\linewidth]{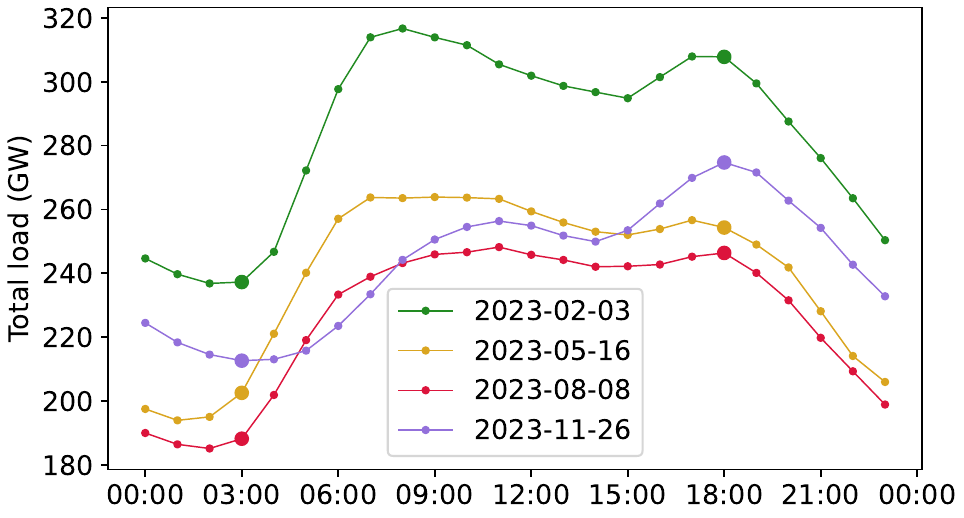}
    \caption{Evolution of the total load for four different days of the year 2023, namely February 3rd, May 16th, August 8th, and November 26th. Larger markers at 3:00 and 18:00 indicate the selected dispatches for our simulations. Times in UTC.}
    \label{fig:all_dispatches}
\end{figure}

In Figure \ref{fig:all_dispatches}, we show the evolution of the total load for four different dates in 2023. In particular, we chose February 3, May 16, August 8, and November 26 to capture the variability of load levels throughout the year. For each of these dates, we selected dispatch times at 3:00 and 18:00, as these times are close to the daily minimum and maximum load respectively.

To determine the regions R and I where we reduce and increase generation, we pick two generators at random and select all generators within a distance from them. We have done this for two different radii of 200 km and 400 km. 

Most of the times, splitting takes place either in the Pyrenees (34 cases) or the Balkans (25 cases). These are included in Sections \ref{sec:supp_pyrenees} and \ref{sec:supp_balkans}. However, we have also observed 5 cases of splitting in France, 4 in Italy, 2 in Central Europe, 4 that involve a relatively large group of nodes, and the remaining 86 cases involve much smaller groups of nodes or leaf nodes. We will discuss some of these results in Section \ref{sec:supp_otherregions}.

\subsection{Pyrenees} \label{sec:supp_pyrenees}

\begin{figure}[H]
    \centering
    \includegraphics[width=0.17\linewidth]{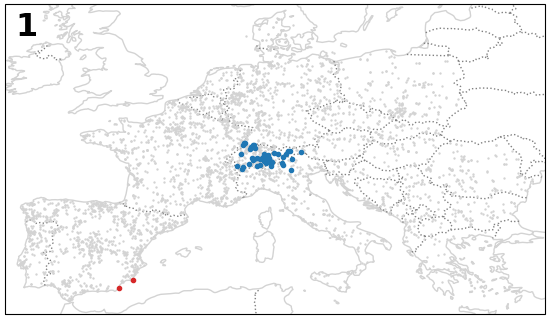}
    \includegraphics[width=0.17\linewidth]{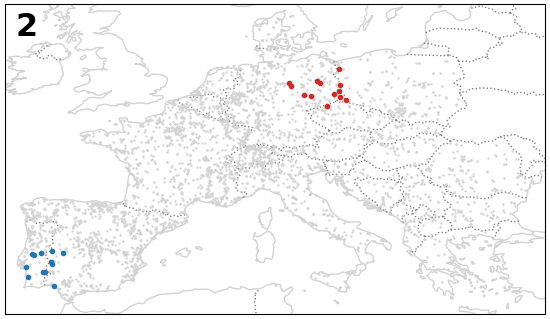}
    \includegraphics[width=0.17\linewidth]{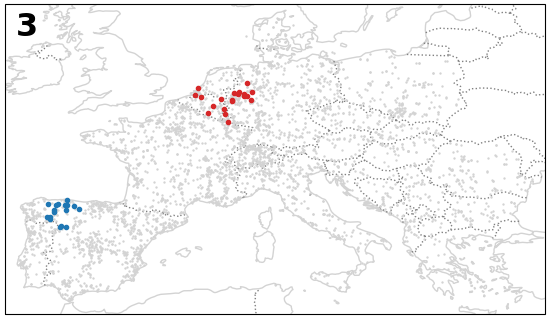}
    \includegraphics[width=0.17\linewidth]{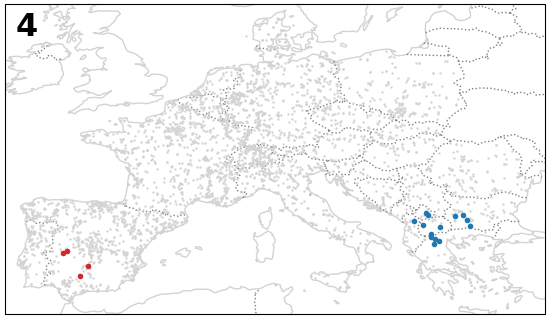}
    \includegraphics[width=0.17\linewidth]{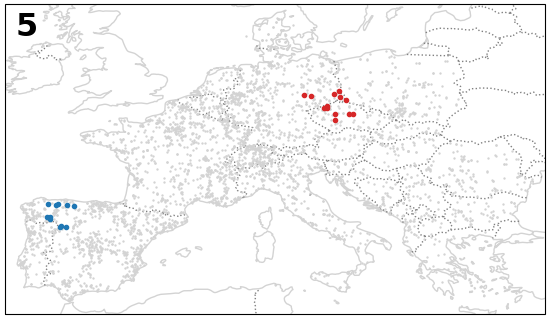}
    \includegraphics[width=0.17\linewidth]{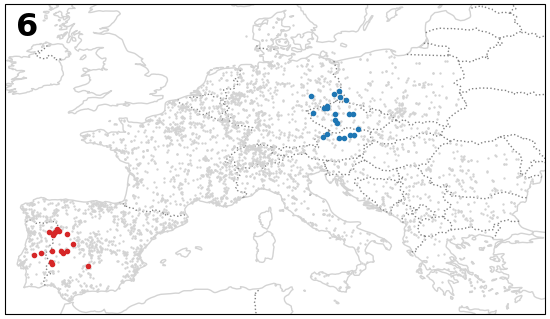}
    \includegraphics[width=0.17\linewidth]{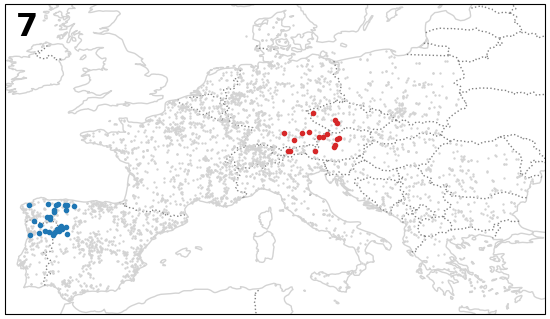}
    \includegraphics[width=0.17\linewidth]{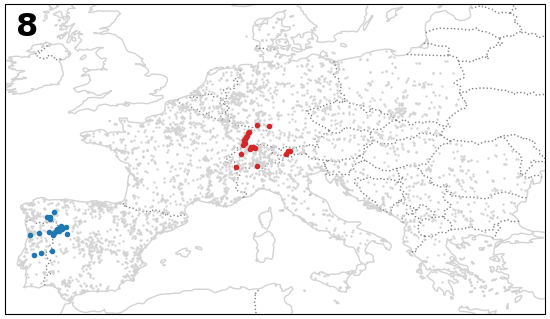}
    \includegraphics[width=0.17\linewidth]{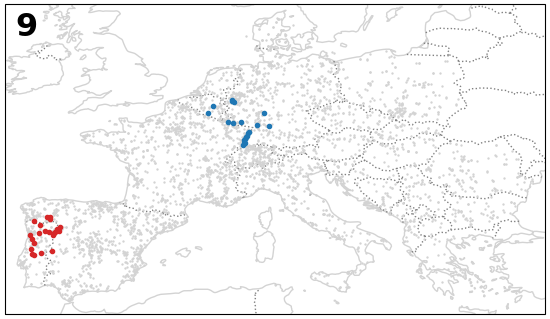}
    \includegraphics[width=0.17\linewidth]{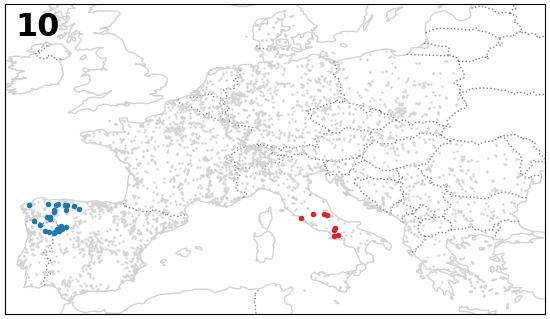}
    \includegraphics[width=0.17\linewidth]{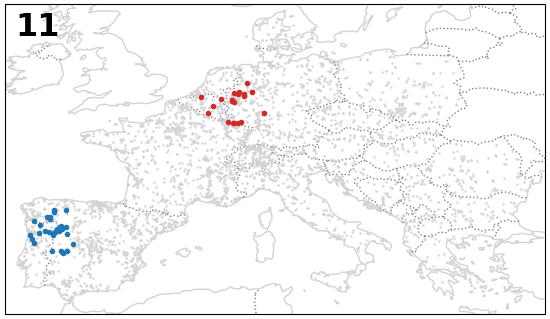}
    \includegraphics[width=0.17\linewidth]{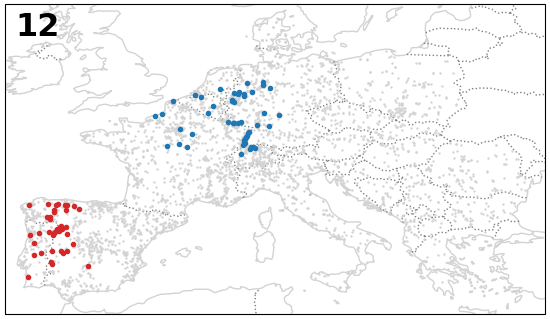}
    \includegraphics[width=0.17\linewidth]{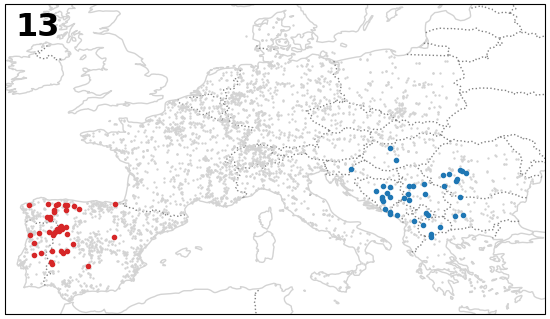}
    \includegraphics[width=0.17\linewidth]{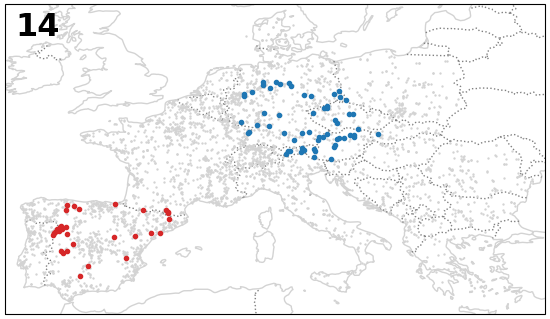}
    \includegraphics[width=0.17\linewidth]{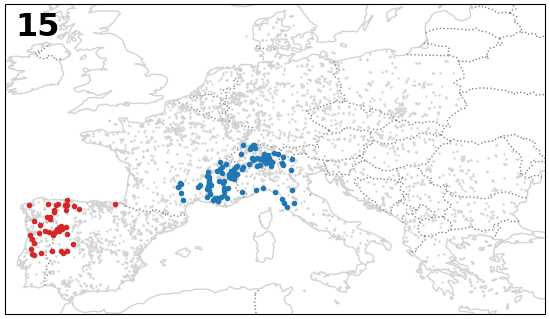}
    \includegraphics[width=0.17\linewidth]{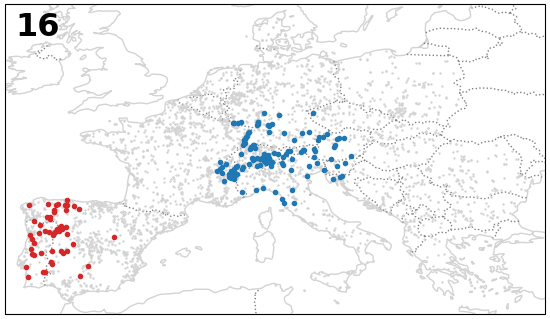}
    \includegraphics[width=0.17\linewidth]{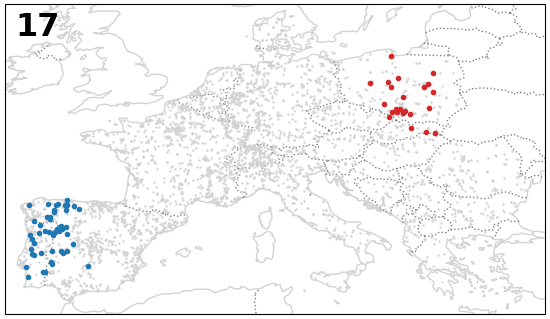}
    \includegraphics[width=0.17\linewidth]{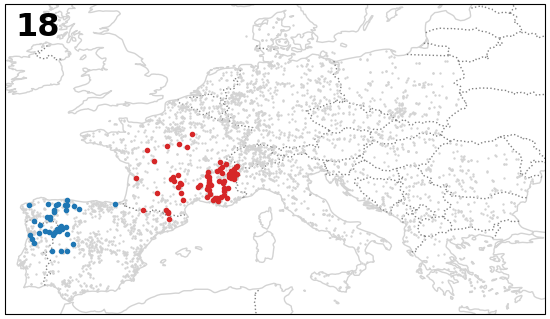}
    \includegraphics[width=0.17\linewidth]{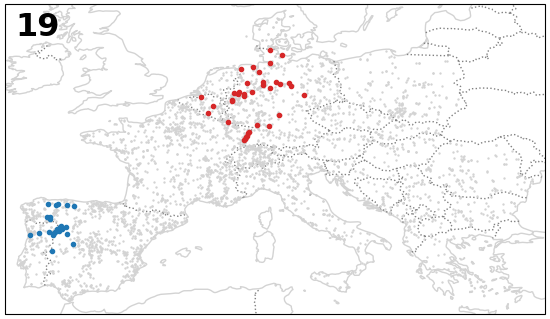}
    \includegraphics[width=0.17\linewidth]{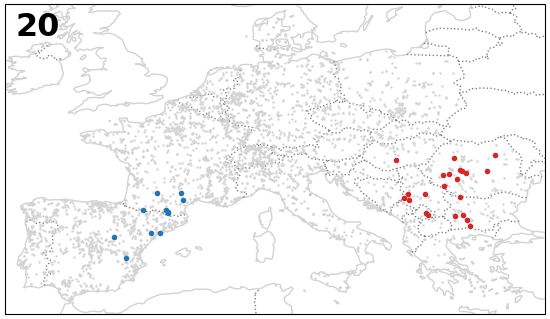}
    \includegraphics[width=0.17\linewidth]{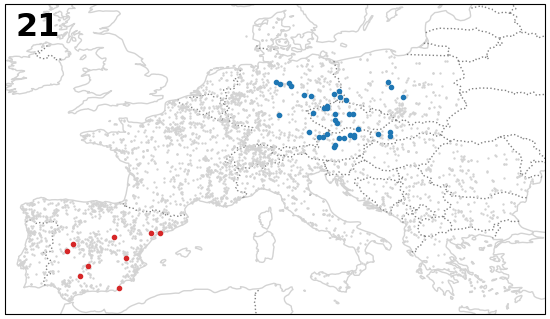}
    \includegraphics[width=0.17\linewidth]{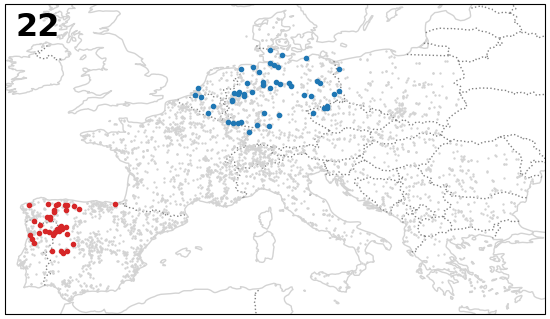}
    \includegraphics[width=0.17\linewidth]{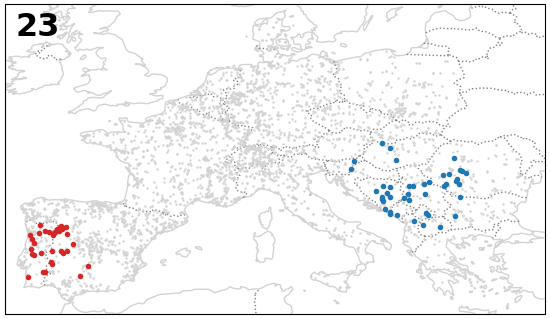}
    \includegraphics[width=0.17\linewidth]{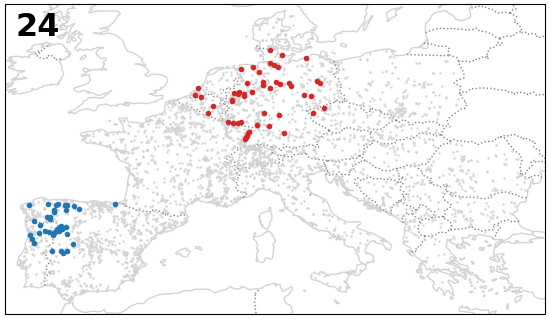}
    \includegraphics[width=0.17\linewidth]{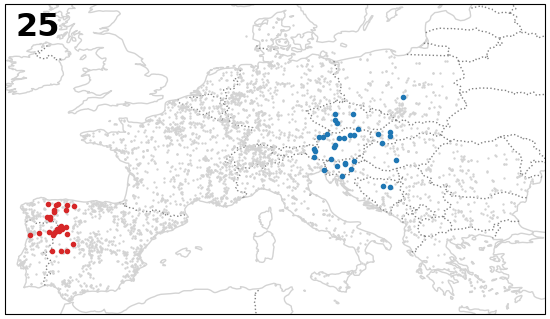}
    \includegraphics[width=0.17\linewidth]{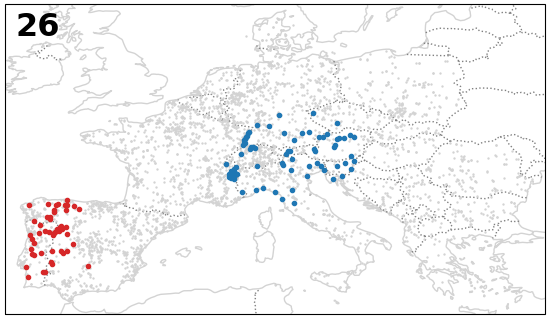}
    \includegraphics[width=0.17\linewidth]{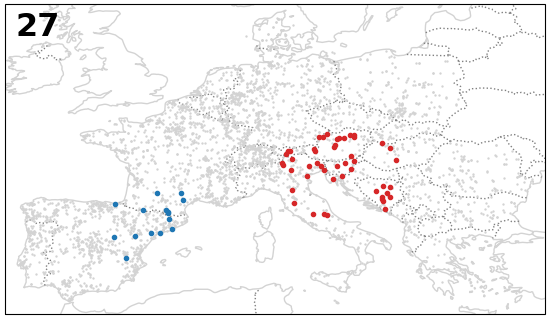}
    \includegraphics[width=0.17\linewidth]{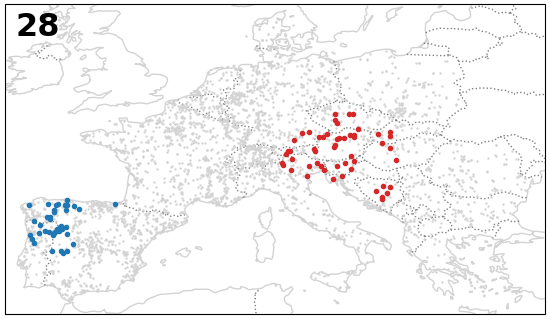}
    \includegraphics[width=0.17\linewidth]{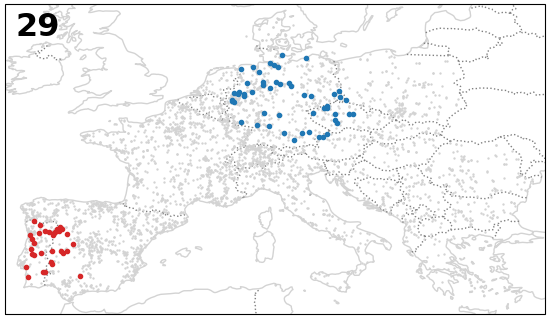}
    \includegraphics[width=0.17\linewidth]{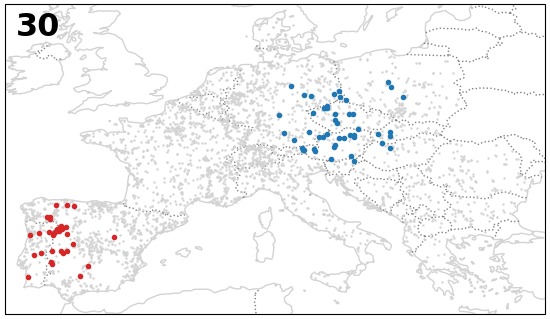}
    \includegraphics[width=0.17\linewidth]{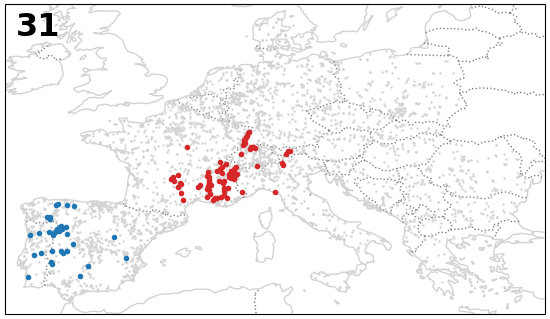}
    \includegraphics[width=0.17\linewidth]{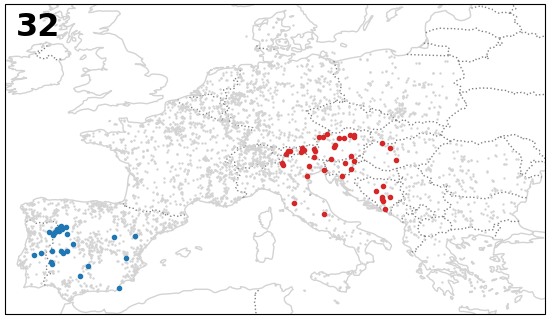}
    \includegraphics[width=0.17\linewidth]{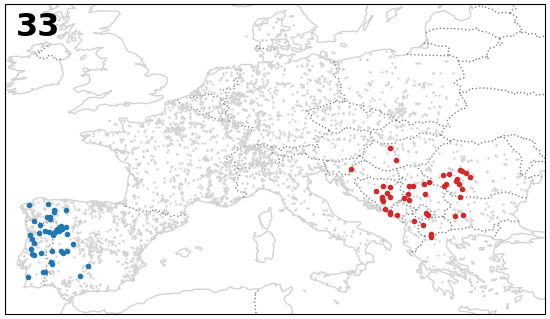}
    \includegraphics[width=0.17\linewidth]{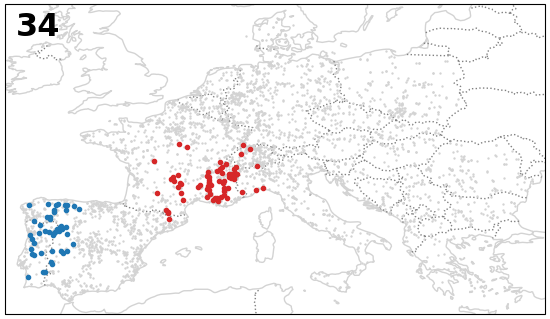}
    \caption{Power imbalance distributions that lead to splitting in the Pyrenees.}
    \label{fig:pyrenees_imbalance}
\end{figure}

In FIG. \ref{fig:pyrenees_imbalance}, we show the analyzed imbalance scenarios that lead to splitting in the Pyrenees. In all these scenarios, the eigenvector associated to the eigenvalue that crosses zero is very similar to that shown in panels (a) and (b) of FIG. \ref{fig:northsouth_centraleast_eigvec} and FIG. \ref{fig:northsouth_centraleast_eigvec_DC}. Moreover, the instability always occurs when the four lines of FIG. \ref{fig:northsouth_centraleast_fixedpoint_DC}(b) are converted to DC.

\subsection{Balkans} \label{sec:supp_balkans}

\begin{figure}[H]
    \centering
    \includegraphics[width=0.17\linewidth]{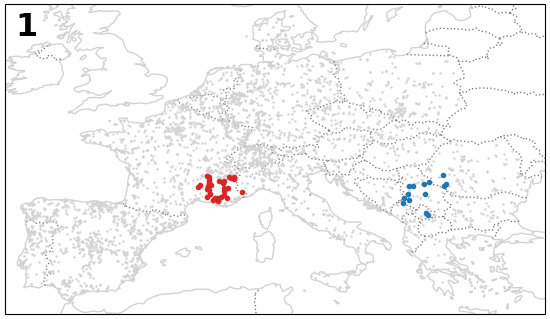}
    \includegraphics[width=0.17\linewidth]{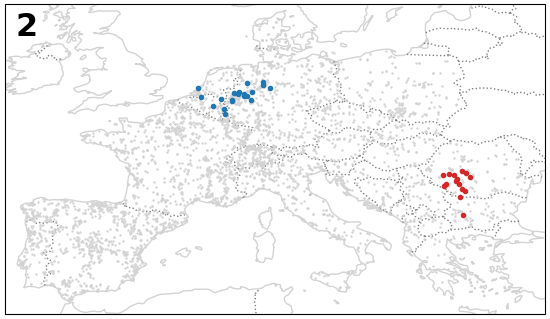}
    \includegraphics[width=0.17\linewidth]{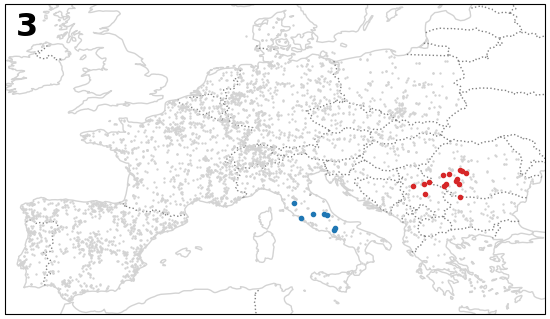}
    \includegraphics[width=0.17\linewidth]{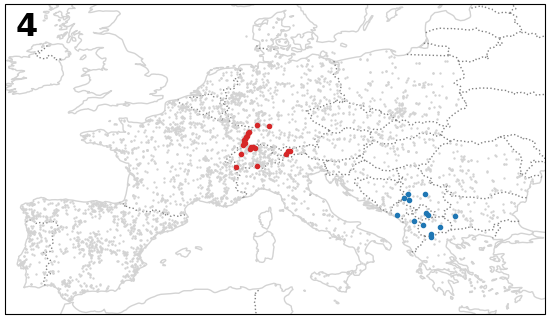}
    \includegraphics[width=0.17\linewidth]{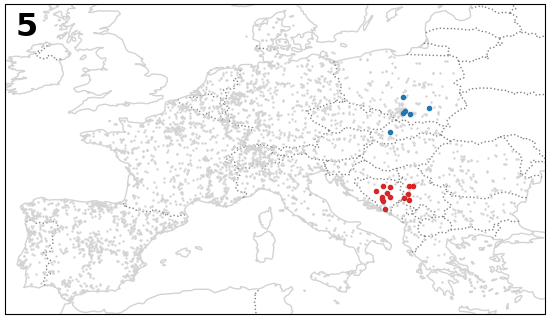}
    \includegraphics[width=0.17\linewidth]{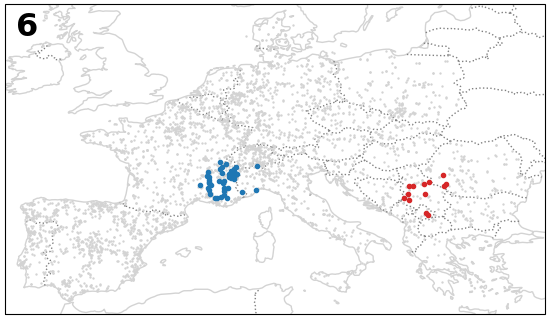}
    \includegraphics[width=0.17\linewidth]{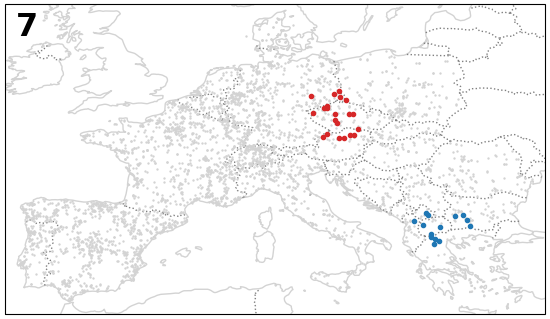}
    \includegraphics[width=0.17\linewidth]{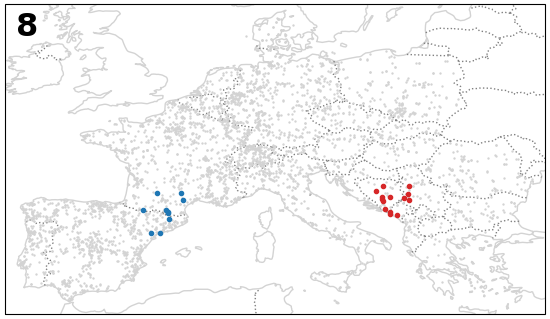}
    \includegraphics[width=0.17\linewidth]{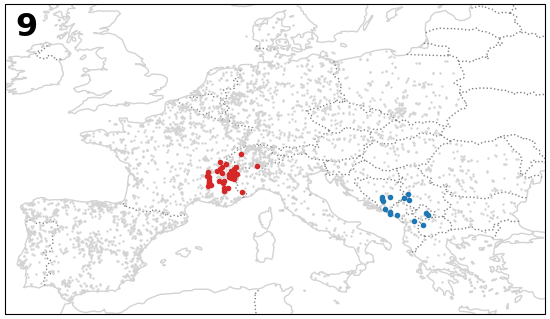}
    \includegraphics[width=0.17\linewidth]{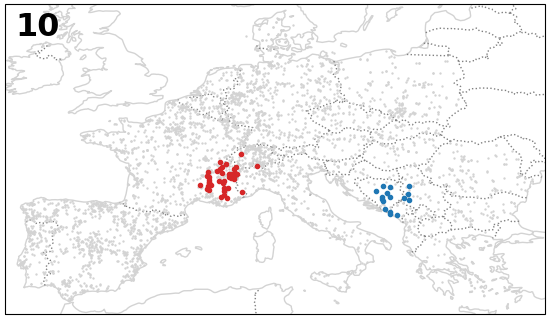}
    \includegraphics[width=0.17\linewidth]{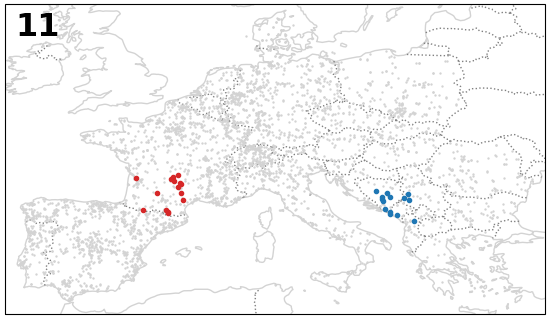}
    \includegraphics[width=0.17\linewidth]{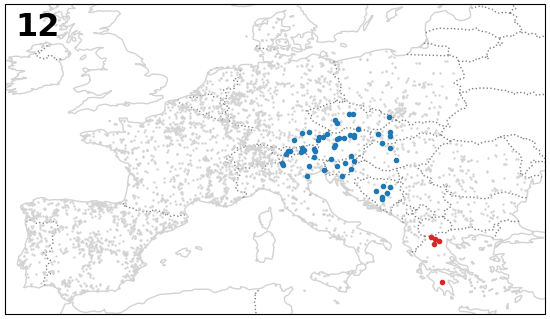}
    \includegraphics[width=0.17\linewidth]{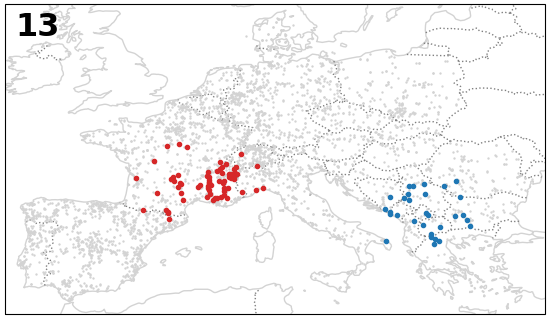}
    \includegraphics[width=0.17\linewidth]{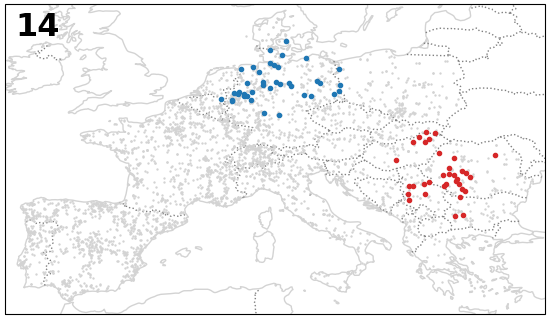}
    \includegraphics[width=0.17\linewidth]{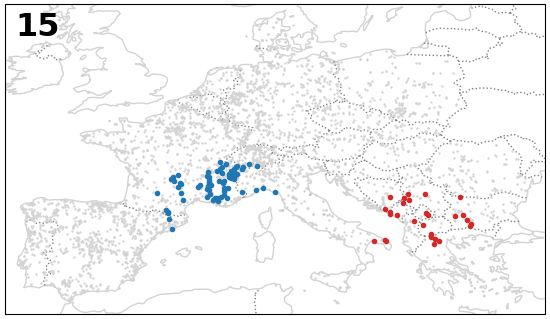}
    \includegraphics[width=0.17\linewidth]{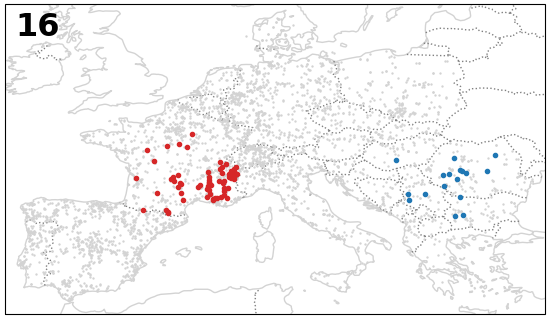}
    \includegraphics[width=0.17\linewidth]{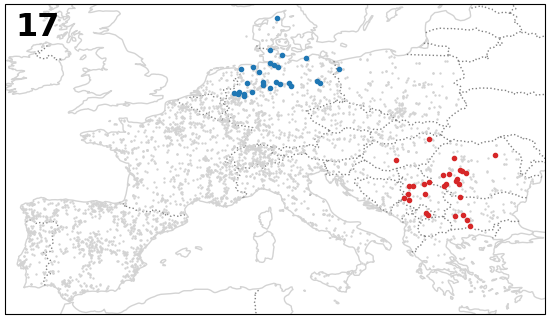}
    \includegraphics[width=0.17\linewidth]{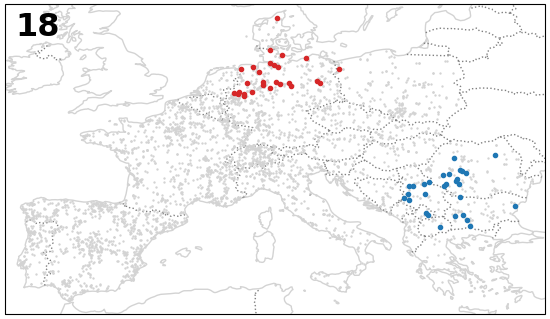}
    \includegraphics[width=0.17\linewidth]{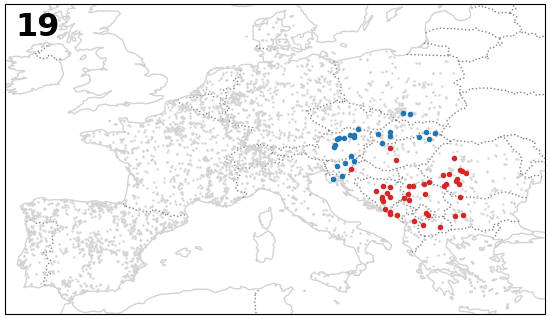}
    \includegraphics[width=0.17\linewidth]{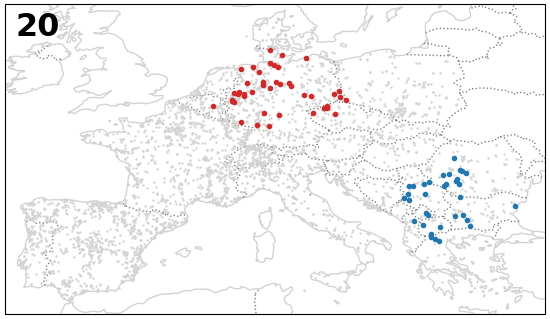}
    \includegraphics[width=0.17\linewidth]{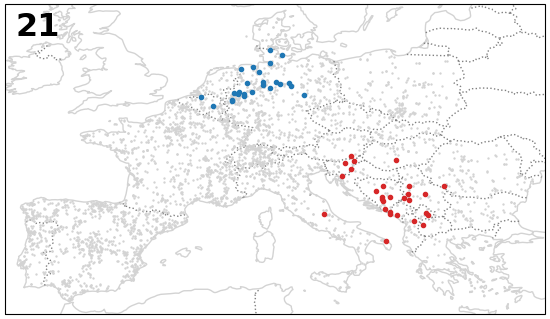}
    \includegraphics[width=0.17\linewidth]{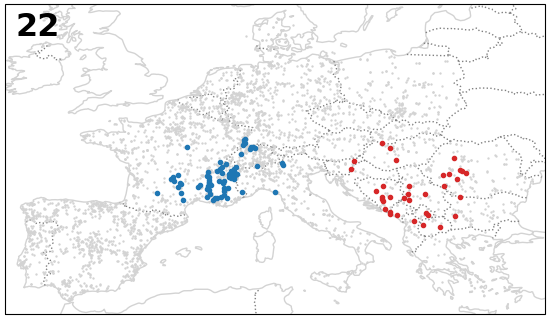}
    \includegraphics[width=0.17\linewidth]{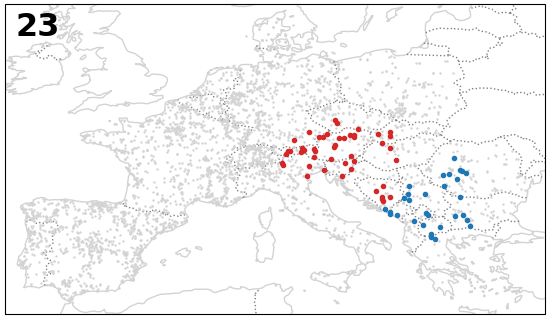}
    \includegraphics[width=0.17\linewidth]{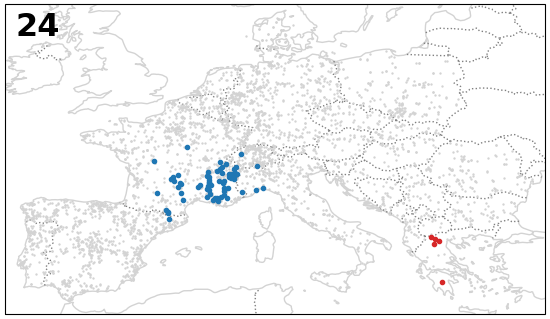}
    \includegraphics[width=0.17\linewidth]{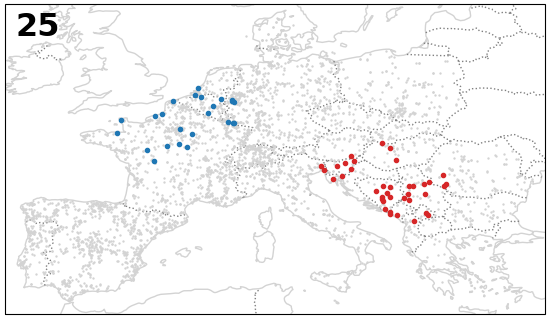}
    \caption{Power imbalance distributions that lead to splitting in the Balkans.}
    \label{fig:balkans_imbalance}
\end{figure}

In FIG. \ref{fig:balkans_imbalance}, we show the analyzed imbalance scenarios that lead to splitting in the Balkans. As for the Pyrenees case, the eigenvector associated to the eigenvalue that crosses zero in all these cases is very similar to that shown in panels (c) and (d) of FIG. \ref{fig:northsouth_centraleast_eigvec} and FIG. \ref{fig:northsouth_centraleast_eigvec_DC}. Moreover, the set of eight lines shown in FIG. \ref{fig:northsouth_centraleast_eigvec_DC}(d) is always involved in the splitting.

\subsection{Other splitting areas} \label{sec:supp_otherregions}

As mentioned in Section \ref{sec:conclusions}, splitting may also take place in other grid areas. Some cases are obvious, such as splitting in Italy, but others are more surprising, as they take place in highly meshed grid areas.

\begin{figure}[H]
    \centering
    \includegraphics[width=.8\linewidth]{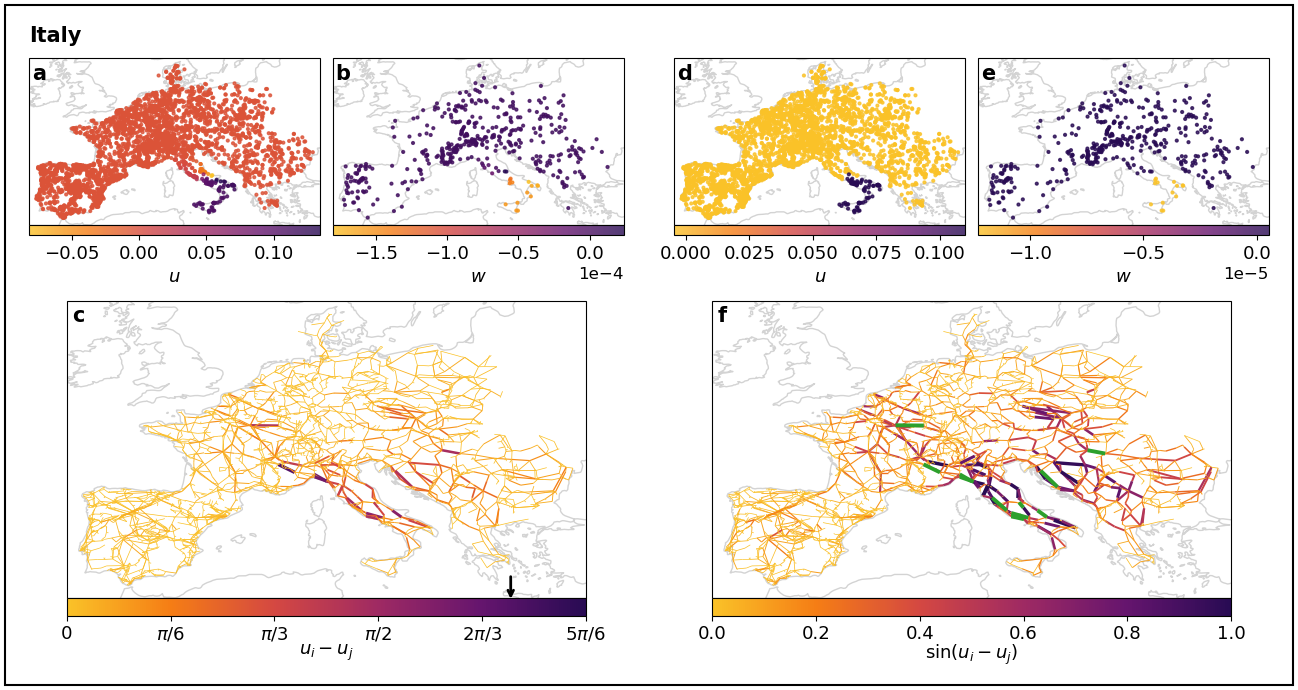}
    \includegraphics[width=.8\linewidth]{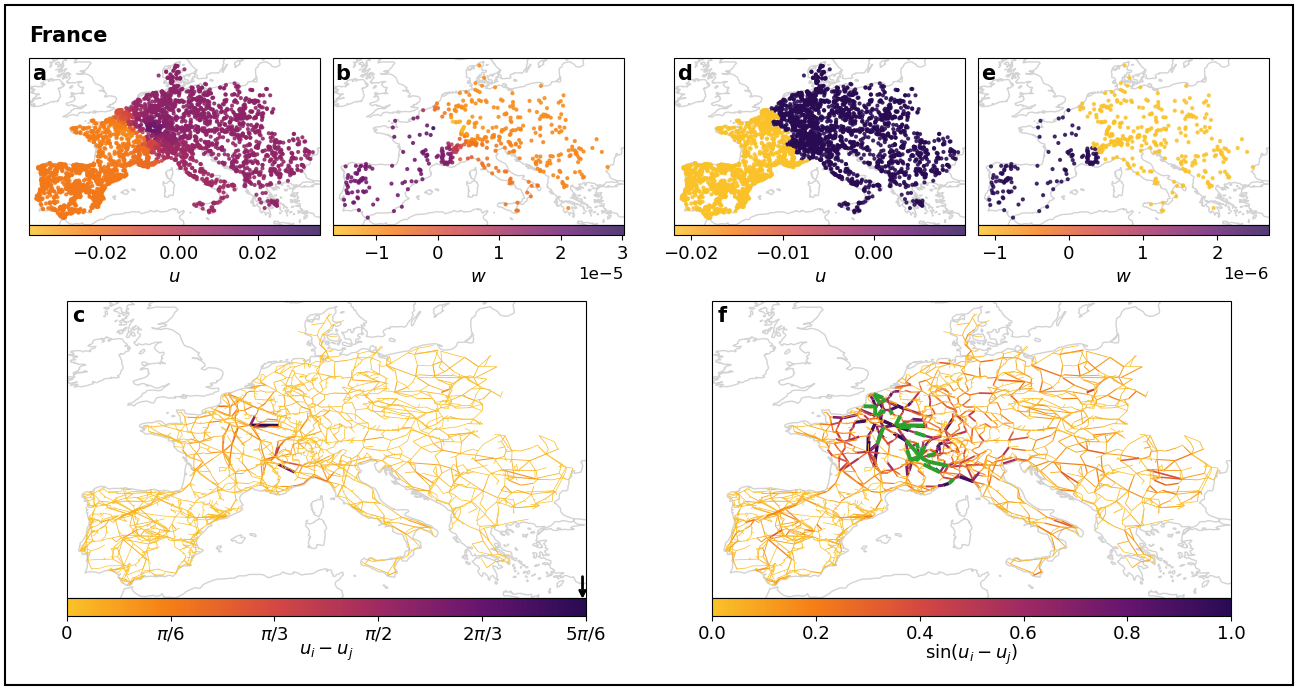}
    \includegraphics[width=.8\linewidth]{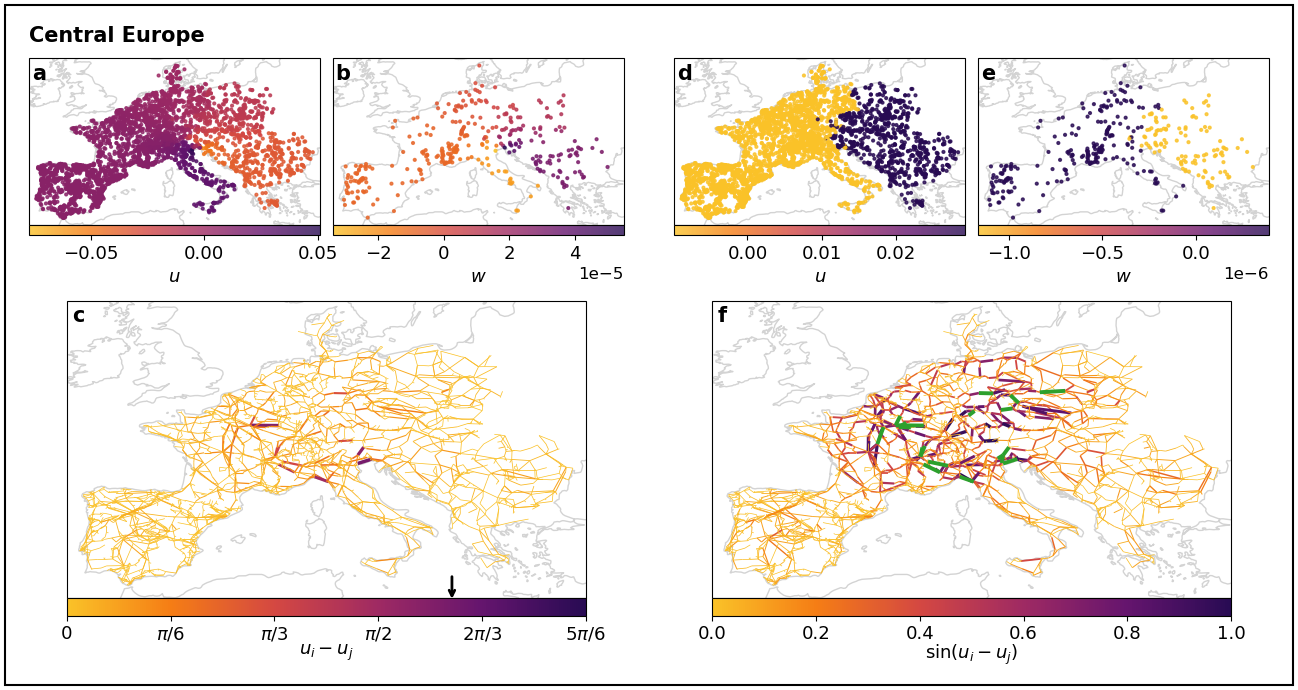}
    \caption{(a,b,c) Initial instability and (d,e,f) after using the DC trick in the imbalance scenarios shown in FIG. \ref{fig:italy_france_centraleurope_imbalances}. (a,d) Phase and (b,e) frequency components of the eigenvector associated with the eigenvalue that reaches zero at the SN. (c) Line phase differences and (f) line stress, where DC lines are shown in green. }
    \label{fig:italy_france_centraleurope_splitting}
\end{figure}

\begin{figure}[H]
    \centering
    \includegraphics[width=.3\linewidth]{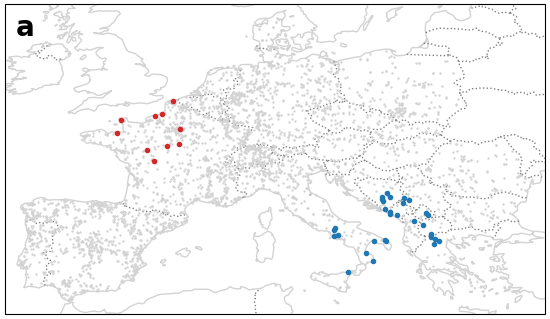}
    \includegraphics[width=.3\linewidth]{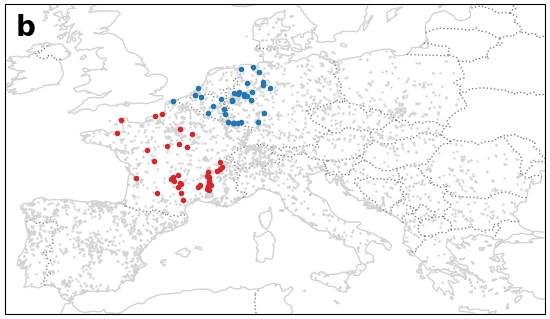}
    \includegraphics[width=.3\linewidth]{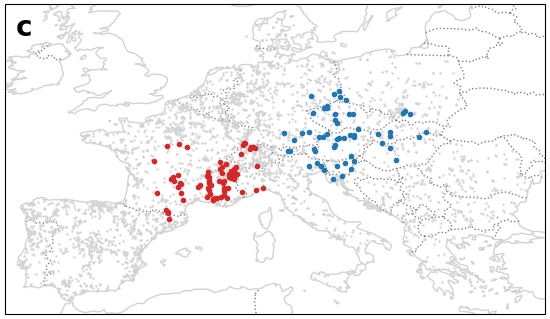}
    \caption{Power imbalances leading to splitting in (a) Italy, (b) France, and (c) Central Europe.}
    \label{fig:italy_france_centraleurope_imbalances}
\end{figure}

In FIG. \ref{fig:italy_france_centraleurope_splitting}, we show three different splitting cases, namely in Italy, France, and Central Europe. These were obtained for the imbalance scenarios shown in FIG. \ref{fig:italy_france_centraleurope_imbalances}.

\vspace{-0.5cm}

